\documentclass[11pt,a4paper]{article} 
\usepackage{style}
\usepackage{amsmath}
\usepackage{amssymb}
\usepackage{amsthm}
\usepackage{psfrag}
\usepackage{graphicx}
\usepackage{hyperref}
\usepackage{feynmp}

\usepackage{dcolumn}

\usepackage[table,svgnames,dvipsnames]{xcolor}
\usepackage{booktabs}

\definecolor{vlightgray}{gray}{0.9}
\usepackage{color}
\usepackage{bm}

\renewcommand{\a}{\alpha}

\newcommand{\be}{\begin{equation}}
\newcommand{\ee}{\end{equation}}
\newcommand{\beqa}{\begin{eqnarray}}
\newcommand{\eeqa}{\end{eqnarray}}
\newcommand{\bsm}{\begin{smallmatrix}}
\newcommand{\esm}{\end{smallmatrix}}

\renewcommand\a{\alpha}

\newcommand{\vp}{\varphi}

\renewcommand\k{{\bf k}}

\newcommand{\hMpc}{h\text{Mpc}^{-1}}

\newcommand{\bseq}{\begin{subequations}}
\newcommand{\eseq}{\end{subequations}}

\renewcommand{\ln}{\mathop{\rm ln}\nolimits}

\renewcommand{\k}{{\bf k}}

\newcommand\vk{\varkappa}

\newcommand{\z}{{\bf z}}
\newcommand{\kmax}{k_{\rm max}}
\newcommand{\kmin}{k_{\rm min}}

\newcommand{\astfootnote}[1]{%
\let\oldthefootnote=\thefootnote%
\setcounter{footnote}{0}%
\renewcommand{\thefootnote}{\fnsymbol{footnote}}%
\footnote{#1}%
\let\thefootnote=\oldthefootnote%
}

\usepackage{color}
\usepackage[caption=false]{subfig}

\renewcommand{\vec}[1]{\boldsymbol{#1}}

\renewcommand{\hMpc}{h\,\mathrm{Mpc}^{-1}}
\newcommand{\delD}[1]{(2\pi)^3\delta_\mathrm{D}\left({#1}\right)}
\newcommand{\av}[1]{\left\langle{#1}\right\rangle} 
\renewcommand{\vk}{\vec k}
\renewcommand{\vp}{\vec p}
\newcommand{\vd}{\vec d}

\newcommand{\vx}{\vec x}

\newcommand{\vy}{\vec y}
\newcommand{\vr}{\vec r}

\newcommand{\ft}[1]{\mathcal{F}\left[{#1}\right]}
\newcommand{\ift}[1]{\mathcal{F}^{-1}\left[{#1}\right]}

\newcommand{\hr}{\hat{\vec r}}
\newcommand{\hn}{\hat{\vec n}}

\newcommand{\hk}{\hat{\vec k}}

\def\beq{\begin{eqnarray}}
\def\eeq{\end{eqnarray}}

\definecolor{darkgreen}{RGB}{0,120,0}

\newcommand{\va}{\vec a}
\newcommand{\Hi}{\mathsf{H}^{-1}}
\newcommand{\Ai}{\mathsf{A}^{-1}}
\newcommand{\B}{\mathsf{B}}

\relax

\title{Cosmology with the Galaxy Bispectrum Multipoles:\\
\Large Optimal Estimation and Application to BOSS Data}

\author[a,b]{Mikhail M. Ivanov,\footnote{\href{mailto:ivanov@ias.edu}{ivanov@ias.edu}}}
\author[c,d,e,a]{Oliver H.\,E. Philcox,\footnote{\href{mailto:ohep2@cantab.ac.uk}{ohep2@cantab.ac.uk}}}
\author[a]{Giovanni Cabass,}
\author[f,g]{Takahiro Nishimichi,} 
\author[h]{Marko Simonovi\'c,}
\author[a]{and Matias Zaldarriaga.}

\affiliation[a]{School of Natural Sciences, Institute for Advanced Study,\\1 Einstein Drive, Princeton, NJ 08540, USA}
\affiliation[b]{NASA Hubble Fellowship Program Einstein Postdoctoral Fellow}%

\affiliation[c]{Center for Theoretical Physics, Department of Physics, Columbia University,\\538 West 120th Street, New York, NY 10027, USA}
\affiliation[d]{Simons Society of Fellows, Simons Foundation, New York, NY 10010, USA}
\affiliation[e]{Department of Astrophysical Sciences, Princeton University,\\Peyton Hall, 4 Ivy Lane, Princeton, NJ 08544, USA}

\affiliation[f]{Center for Gravitational Physics and Quantum Information, \\ Yukawa Institute for Theoretical Physics, Kyoto University, Kyoto 606-8502, Japan}
\affiliation[g]{Kavli Institute for the Physics and Mathematics of the Universe (WPI), UTIAS \\The University of Tokyo, Kashiwa, Chiba 277-8583, Japan}

\affiliation[h]{Theoretical Physics Department, CERN,\\1 Esplanade des Particules, Geneva 23, CH-1211, Switzerland}

\abstract{\small
We present a framework for self-consistent cosmological analyses of the full-shape anisotropic bispectrum, including the quadrupole $(\ell=2)$ and hexadecapole $(\ell=4)$ moments. This features a novel window-free algorithm for extracting the latter quantities from data, derived using a maximum-likelihood prescription. Furthermore, we introduce a theoretical model for the bispectrum multipoles (which does not introduce new free parameters), and test both aspects of the pipeline on several high-fidelity mocks, including the PT Challenge suite of gigantic cumulative volume. This establishes that the systematic error is significantly below the statistical threshold, both for the measurement and modeling. As a realistic example, we extract the large-scale bispectrum multipoles from BOSS DR12 and analyze them in combination with the power spectrum data. Assuming a minimal $\Lambda$CDM model, with a BBN prior on the baryon density and a \textit{Planck} prior on $n_s$, we can extract the remaining cosmological parameters directly from the clustering data. The inclusion of the unwindowed higher-order $(\ell>0)$ large-scale bispectrum multipoles is found to moderately improve one-dimensional cosmological parameter posteriors (at the $5\%-10\%$ level), though these multipoles are detected only in three out of four BOSS data segments at $\approx 5\sigma$.
Combining information from the power spectrum and bispectrum multipoles, the real space power spectrum, and the post-reconstructed BAO data, we find $H_0 = 68.2\pm 0.8~\mathrm{km}\,\mathrm{s}^{-1}\mathrm{Mpc}^{-1}$, $\Omega_m =0.33\pm 0.01$ and $\sigma_8 = 0.736\pm 0.033$ (the tightest yet found in perturbative full-shape analyses). Our estimate of the growth parameter $S_8=0.77\pm 0.04$ agrees with both weak lensing and CMB results. The estimators and data used in this work have been made \href{https://github.com/oliverphilcox/Spectra-Without-Windows}{publicly available}.
}
\normalsize

\begin{document}

\begin{flushright}
YITP-23-13, CERN-TH-2023-022
\end{flushright}

\maketitle

\section{Introduction}

The large scale structure (LSS) traced by the distribution of galaxies, has become one of the primary cosmological observables, allowing for precision tests of our theoretical models and numerical simulations. A key feature of this distribution is its statistical non-Gaussianity, induced by non-linear gravitational evolution. Any analysis aimed at maximizing the information yield of a galaxy survey should therefore include non-Gaussian statistics, the simplest of which is the three-point correlation function of the galaxy overdensity field, or its Fourier image, known as  the bispectrum. 

Spectroscopic surveys observe the galaxy distribution in three dimensions, with the radial axis contaminated by line-of-sight velocities, through the phenomena of redshift space distortions (RSD). This anisotropy propagates to summary statistics such as the bispectrum \cite{Scoccimarro:1999ed,Scoccimarro:2000sn}, and is a valuable probe of cosmological information encoded in the peculiar velocity field. To date, most bispectrum analyses to date have considered only the angle-averaged galaxy bispectrum, also called the bispectrum monopole moment \citep[e.g.,][]{1975ApJ...196....1P,1977ApJ...217..385G,Feldman:2000vk,Marin:2013bbb,Scoccimarro:2000sp,Scoccimarro:1997st,Sefusatti:2006pa,Gil-Marin:2014sta,Gil-Marin:2014baa,Kitaura:2014mja,Hahn:2019zob,Chudaykin:2019ock,Ivanov:2021kcd,Philcox:2021kcw}. This moment, however, is only the first term of an infinite expansion in angular moments needed to capture the entire anisotropic clustering information present within the bispectrum \citep{Scoccimarro:1999ed,Scoccimarro:2015bla}. Including this information in analysis pipelines requires a systematic and efficient treatment, taking careful account of effects such as analytical modeling, robust statistical estimation, the impact of survey geometry, and discreteness effects. In this work, we present the first such analysis carried out on publicly available data using the twelfth data release of the Baryon Oscillation Spectroscopic Survey (BOSS)~\cite{Alam:2016hwk}.

A number of previous works have studied the galaxy bispectrum beyond the monopole moment including Refs.\,\cite{Yankelevich:2018uaz,Bharadwaj:2020wkc,Mazumdar:2020bkm,Gualdi:2020ymf,Agarwal:2020lov,Mazumdar:2022ynd,Rizzo:2022lmh,DAmico:2022osl,Tsedrik:2022cri} (see also Refs.\,\cite{Sefusatti:2007ih,Sefusatti:2009qh,Baldauf:2016sjb,Hahn:2019zob,Welling:2016dng,MoradinezhadDizgah:2018ssw,Hahn:2020lou,MoradinezhadDizgah:2020whw,Ruggeri:2017dda,Song:2015gca,Karagiannis:2018jdt,Peloso:2013zw,Kehagias:2013yd,Valageas:2013cma,Creminelli:2013mca,Creminelli:2013poa,Creminelli:2013nua,Lewandowski:2019txi,Crisostomi:2019vhj,Oddo:2019run,Oddo:2021iwq,Chudaykin:2019ock} for other bispectrum analyses). Using a combination of Fisher forecasts and simulated data, several of these works have demonstrated that anisotropic bispectrum multipoles may lead to a significant tightening of our constraints on cosmological and astrophysical parameters of interest; for example, Ref.\,\citep{Rizzo:2022lmh} studied the information content in the idealized setting of periodic box geometries with tree-level perturbation theory and derived cosmological parameters such as $f\sigma_8(z)$. Here, our goal is to extend these studies by considering their application both to actual data (including all relevant observational effects and covariances) and to the measurement of underlying $\Lambda$CDM cosmology parameters, thus discovering whether the purported gains can be practically realized. An important step towards this was performed in Ref.~\cite{DAmico:2022osl}, which analyzes observational data from the BOSS bispectrum quadrupole, using tree-level theory. This work finds more modest improvements from the redshift-space information, with only a small ($<10\%$) posterior shrinkage observed for $\omega_{\rm cdm}$ (and $\Omega_m$). Here, we go beyond the former work by including a more detailed treatment of survey geometry effects (\textit{i.e.}\ window-function convolution), and through testing the pipeline on high-quality large-volume simulations, ensuring that our results remain applicable to future high-precision surveys.

Here, our goal is to perform a systematic, consistent, and efficient analysis of the large-scale galaxy bispectrum quadrupole and hexadecapole, as applied to realistic survey data. In this vein, we will address several key issues that have previously complicated anisotropic galaxy bispectrum analyses. First, we validate our perturbative theoretical model for the bispectrum multipoles (based on \cite{Ivanov:2021kcd}) on the high-fidelity PT Challenge simulation dataset~\cite{Nishimichi:2020tvu}. This allows us to test our fitting pipeline in the unprecedented conditions that correspond to the cumulative volume of $566h^{-3}\mathrm{Gpc}^3$, which significantly exceeds the volume of upcoming and even futuristic surveys. 

To robustly account for the mixing of modes and multipoles induced by the survey geometry, we will construct new `window-free' estimators for the bispectrum multipoles, based on the maximum-likelihood approaches outlined in \citep{Philcox:2020vbm,Philcox:2021ukg}.\footnote{A public implementation of these is available at \href{https://github.com/oliverphilcox/Spectra-Without-Windows}{GitHub.com/OliverPhilcox/Spectra-Without-Windows}.} This approach is tested using a suite of Nseries mocks, designed for precision tests of the official BOSS analysis pipeline~\cite{BOSS:2016wmc}. Our new window-free estimator enables straightforward comparison of theory and data the need to forward model the effect of the window function on the former \cite{Gil-Marin:2014baa}. {This allows us to avoid making simplified assumptions about the window function's action, which have led to the excision of large-scale modes in \cite{DAmico:2022osl}; this could severely limit analyses of primordial non-Gaussianity. Whilst analytic methods for bispectrum convolution now exist} (at least for the monopole, see \citep[e.g.,][]{Pardede:2022udo,Alkhanishvili:2022sov} for recent progress), this route still leads to a significant amplification in model complexity, which may make typical Monte Carlo Markov Chain (MCMC) analyses (with~$\sim 10^6$ steps~\cite{Chudaykin:2020hbf}) infeasible. Our efforts herein are a natural extension of our previous full-shape BOSS analyses of the galaxy power spectrum~\cite{Ivanov:2019pdj,Ivanov:2019hqk,Chudaykin:2020ghx}, BAO~\cite{Philcox:2020vvt}, real-space power spectrum proxy \citep{Ivanov:2021fbu}, and bispectrum monopole \cite{Philcox:2021kcw,Cabass:2022wjy,Cabass:2022ymb}, based on the effective field theory of large-scale structure (EFTofLSS; \cite{Baumann:2010tm,Carrasco:2012cv,Cabass:2022avo,Ivanov:2022mrd}). Alternative BOSS full-shape analyses have been carried out in Refs.~\cite{Lange:2021zre,Chen:2021wdi,Zhang:2021yna,Kobayashi:2021oud,Chen:2022jzq,Lange:2023khv,DAmico:2020kxu,DAmico:2019fhj,DAmico:2020tty,DAmico:2022gki,DAmico:2022osl}. Throughout this work, we focus on the bispectrum multipoles on large scales, \textit{i.e.}\ considering only modes with $k<0.08~\hMpc$. For this reasons we use only the tree-level bispectrum likelihood, though extensions to higher $k$ with the one-loop theory of \citep{Philcox:2022frc} may prove interesting. 

Having extensively tested our pipeline on various mock data, we apply it to the BOSS~DR12 anisotropic clustering measurements. Our overall conclusion is that the BOSS bispectrum multipoles do not carry a significant signal, but their inclusion in the analysis allows one to slightly improve constraints on cosmological parameters. In particular, using priors on the primordial power spectrum tilt $n_s$ from \textit{Planck} 2018 \citep{Aghanim:2018eyx} and a BBN prior on the physical baryon density $\omega_b$, we find the Hubble constant $H_0 = 68.2\pm 0.8~\mathrm{km}\,\mathrm{s}^{-1}\mathrm{Mpc}^{-1}$, the matter density fraction $\Omega_m =0.33\pm 0.01$ and the late-time mass clustering amplitude $\sigma_8 = 0.736\pm 0.033$. The latter two measurements can be combined into a growth parameter $S_8\equiv\sigma_8(\Omega_m/0.3)^{0.5}=0.77\pm 0.04$, which agrees well with other independent estimates from the weak lensing and cosmic microwave background radiation surveys.

Our paper is structured as follows. We begin in \S\ref{sec:summary} by summarizing our main results and placing them in context of other cosmological parameter estimates. In \S\ref{sec: bk-mult} we define the bispectrum multipoles and present idealized estimators before considering their optimal unwindowed form in \S\ref{sec:estim}. Then, \S\ref{sec:theory} reviews our theory model for the redshift-space bispectrum multipoles at the tree-level order in perturbation theory. Our data and likelihood are discussed in detail in \S\ref{sec:likelihood}, and the pipeline validated on mock clustering data from PT Challenge and Nseries simulations in \S\ref{sec:mocks}. Finally, we present our of the BOSS survey data in \S\ref{sec:boss} before concluding with a discussion in \S\ref{sec:disc}.

\begin{figure}
    \centering
    \includegraphics[width=0.55\textwidth]{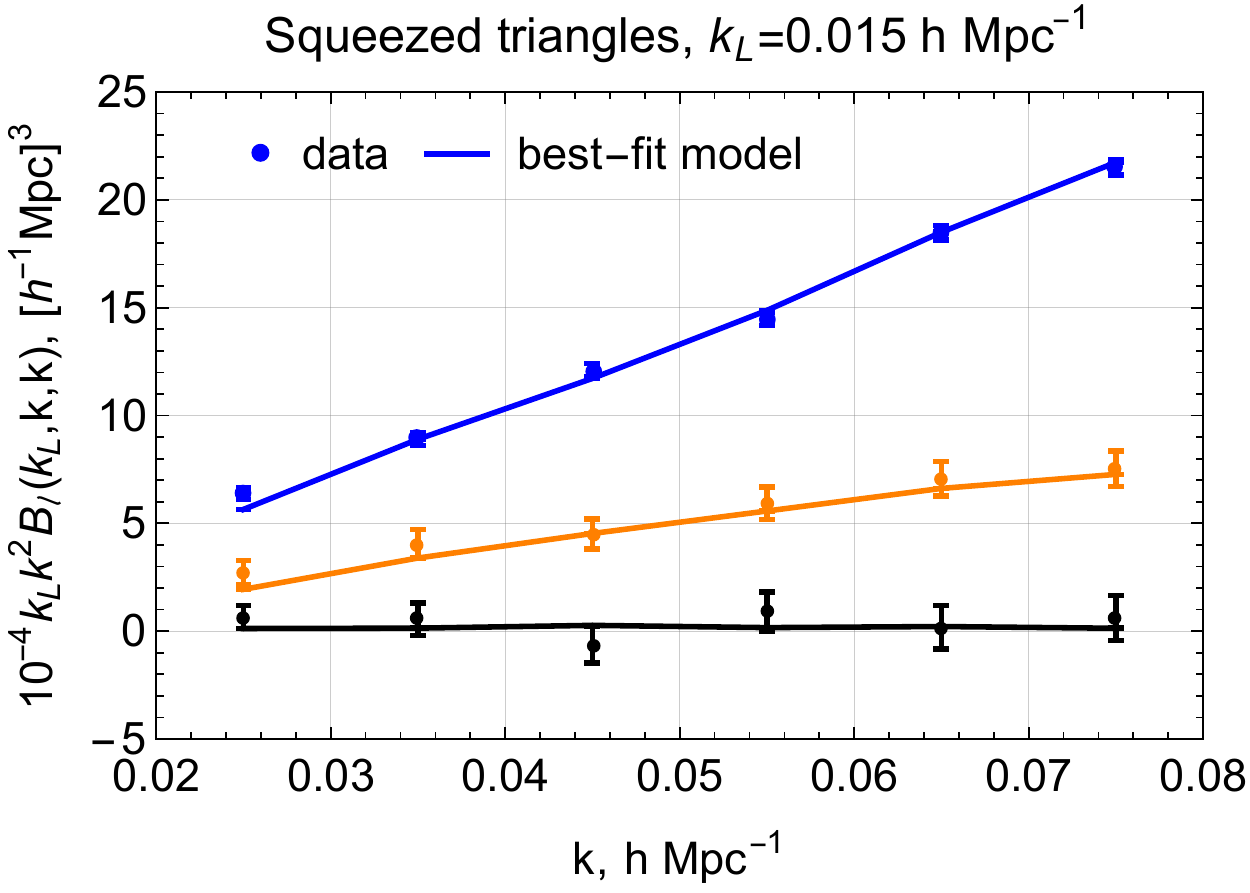}
        \includegraphics[width=0.55\textwidth]{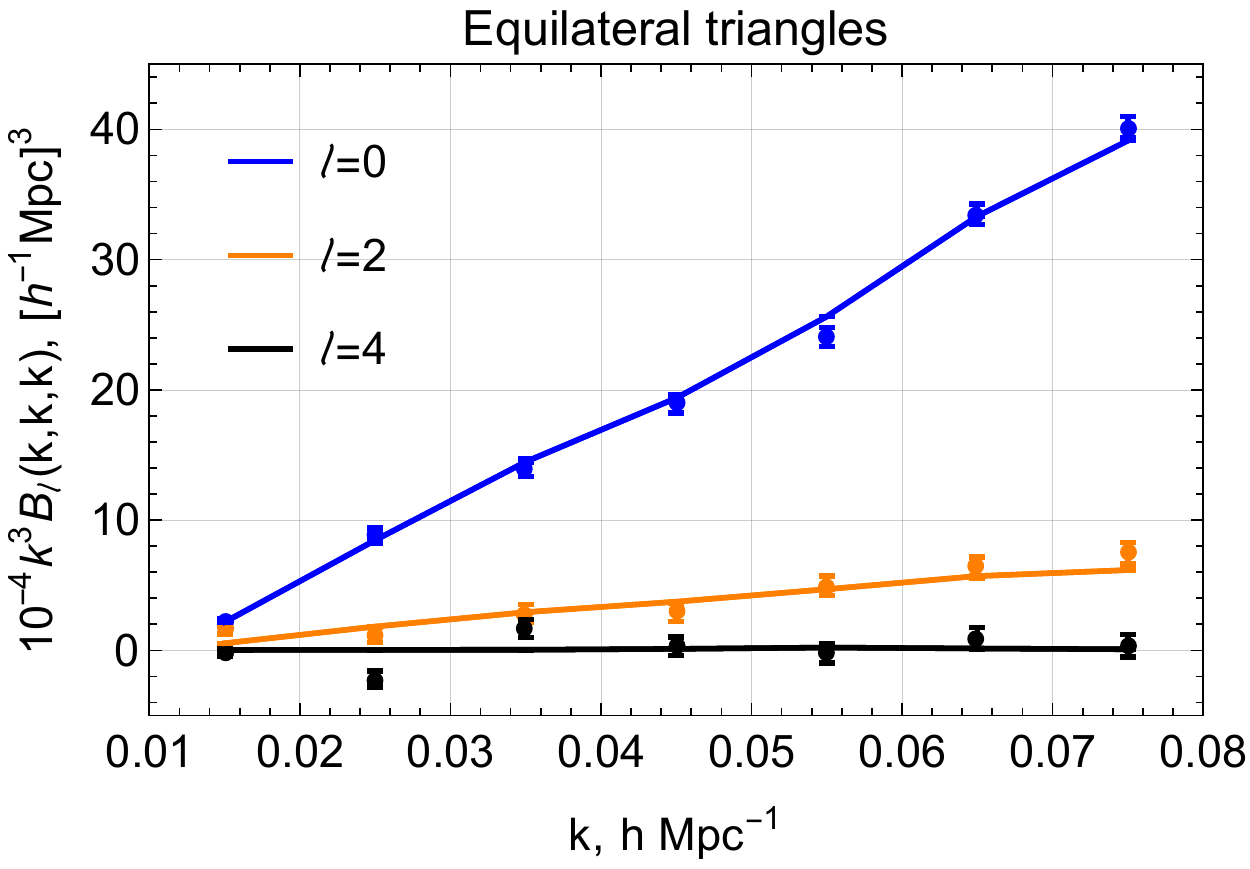}
    \includegraphics[width=0.55\textwidth]{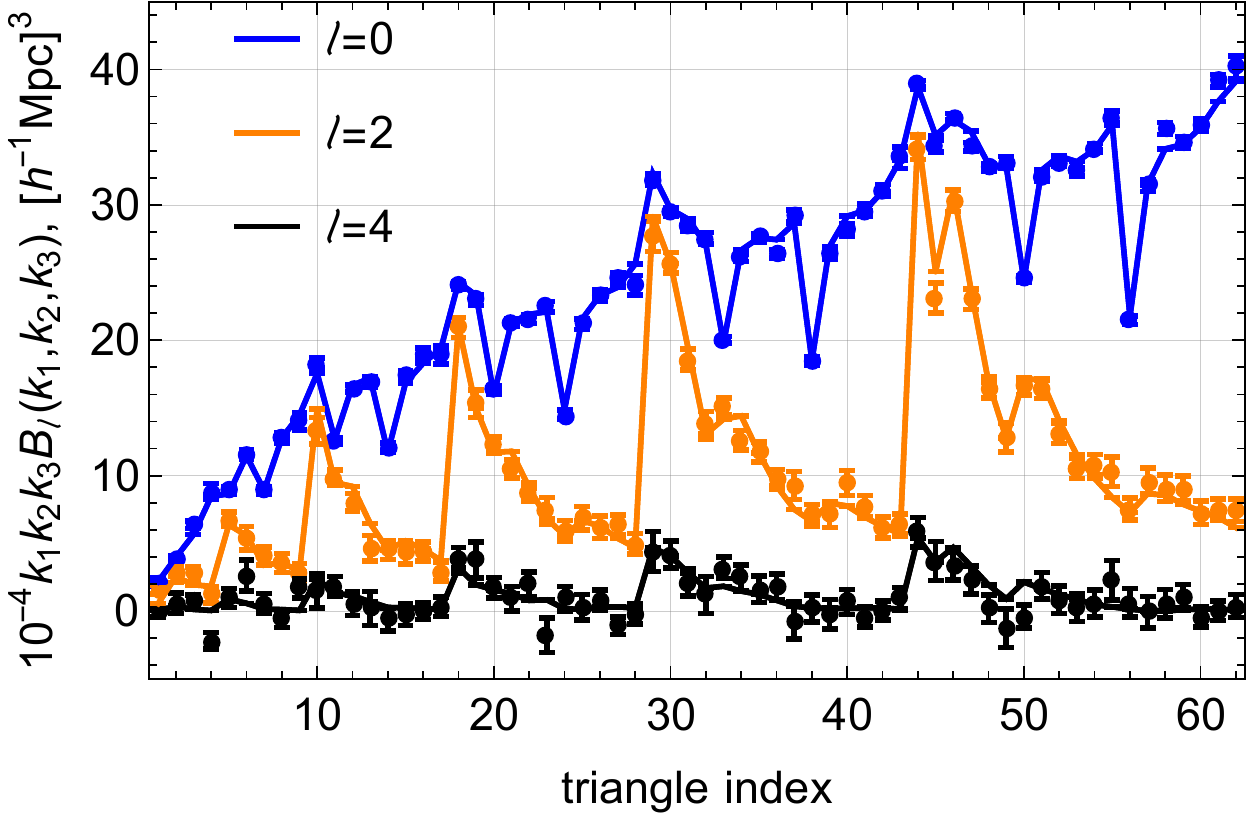}
    \caption{Bispectrum monopole, quadrupole, and hexadecapole extracted from the PT Challenge dataset (points), along with the best-fitting theory model curves (lines). We highlight squeezed and equilateral configurations as a function of wavenumber in the top panels, and show all configurations as a function of the triangle index in the lower panel. The errorbars shown correspond to the diagonal elements of the Gaussian tree-level covariance matrix (see Appendix \ref{sec:gauss}), which matches the total simulation volume of 566~($h^{-1}$Gpc)$^3$. {We note that the extension of the theory model to bispectrum multipoles does not add new parameters. Corresponding detection significances are given in Tab.\,\ref{tab:snrt}.}}
    \label{fig:ptc-plot}
\end{figure}

\section{Summary of the Main Results}
\label{sec:summary}

\begin{table}[!t]
     \centering
     \scriptsize
     \rowcolors{2}{white}{vlightgray}
   \begin{tabular}{c|cccc|ccc} \hline\hline
      \textbf{Dataset} & $\omega_{\rm cdm}$ & $H_0$ & $\ln\left(10^{10}A_{s }\right)$ & $n_s$ & $S_8$ &$\Omega_m$ &$\sigma_8$\\\hline
      $P_\ell+Q_0+B_0$ \,&\, $0.140^{+0.010}_{-0.013}$ \,&\, $69.3\pm 1.1$ \,&\, $2.60\pm 0.13$ \,&\, $0.872\pm 0.066$ \,&\, $0.734\pm 0.039$ \,&\, $0.339^{+0.016}_{-0.018}$ \,&\, $0.691^{+0.035}_{-0.039}$\\
      $P_\ell+Q_0+B_\ell$ \,&\, $0.1444^{+0.0098}_{-0.012}$ \,&\, $69.19^{+0.98}_{-1.1}$ \,&\, $2.60\pm 0.12$\,&\, $0.869\pm 0.060$\,&\,$0.760\pm 0.039$\,&\,$0.349^{+0.015}_{-0.017}$\,&\,$0.704^{+0.034}_{-0.039}$\\\hline
      $P_\ell+Q_0+B_0$\,&\, $0.1262^{+0.0052}_{-0.0058}$\,&\,$68.32\pm 0.83$\,&\,$2.741\pm 0.095$\,&\,--\,&\,$0.745\pm 0.039$\,&\,$0.3197\pm 0.0096$\,&\,$0.722^{+0.032}_{-0.035}$\\
      $P_\ell+Q_0+B_\ell$\,&\, $0.1303\pm 0.0055$\,&\,$68.19\pm 0.78$\,&\,$2.740\pm 0.091$\,&\,--\,&\,$0.771\pm 0.039$\,&\,$0.3296\pm0.0095$\,&\,$0.736\pm 0.033$\\\hline\hline
      \end{tabular}
  \caption{Marginalized constraints on $\Lambda$CDM cosmological parameters from the BOSS power spectrum multipoles, the real-space power spectrum proxy, and the bispectrum. We include BAO information from reconstructed power spectra in all cases. The first and third columns correspond to the likelihood with the bispectrum monopole only, whilst the second and fourth also contain the bispectrum quadrupole and hexadecapole. In each case, we display the mean value and the $68\%$ confidence intervals. All results are obtained assuming the BBN prior on $\omega_b$, with the lower two rows including the \textit{Planck} prior on $n_s$. The final three parameters in each row are derived from the MCMC samples and not sampled directly.}
\label{tab:boss}
\end{table}

\begin{table}[!t]
     \centering
     \scriptsize
     \rowcolors{2}{white}{vlightgray}
   \begin{tabular}{c|cccc} \hline\hline
      \textbf{Dataset} & $B_0$ & $B_2$ & $B_4$ & $B_2+B_4$\\\hline
      BOSS NGC $z=0.61$ \,&\, $390.0~(19.7\sigma)$ \,&\, $24.5~(4.9\sigma)$ \,&\, $2.84~(1.7\sigma)$  \,&\, $23.4~(4.8\sigma)$ \\
      BOSS SGC $z=0.61$ \,&\, $149.4~(12.2\sigma)$ \,&\, $-7.61~(-)$ \,&\, $0.04~(-)$  \,&\, $-7.2~(-)$  \\
      BOSS NGC $z=0.38$\,&\, $271.0~(16.5\sigma)$\,&\,$31.6~(5.6\sigma)$\,&\,$2.7~(1.6\sigma)$ \,&\, $30.2~(5.5\sigma)$ \\
     BOSS SGC $z=0.38$\,&\, $99.7~(10.0\sigma)$\,&\,$15.4~(3.9\sigma)$\,&\,$0.09~(-)$ \,&\, $15.8~(4.0\sigma)$ \\\hline
       PT Challenge $z=0.61$\,&\, $3.07\times 10^5~(554\sigma)$\,&\,$1.88\times 10^4~(137\sigma)$\,&\,$1038~(32\sigma)$ \,&\, $1.90\times 10^4~(138\sigma)$\\\hline\hline
      \end{tabular}
  \caption{$\Delta \chi^2$ values and the
  associated detection significance for the bispectrum multipoles for the four chunks of the BOSS dataset and the PT Challenge simulations. These are computed as $\Delta \chi^2=
  \chi^2_{\rm null}-\chi^2_{\rm model}$, 
  where $\chi^2_{\rm model/null}=\sum_{\ell\ell'}(B^{\rm data}_\ell -B^{\rm model/null}_\ell) \cdot \mathsf{C}^{-1}_{\ell \ell'}\cdot (B^{\rm data}_{\ell'}-B^{\rm model/null}_\ell)$, with $B^{\rm model}_\ell=\{B^{\rm bf}_0,0,0\}$ and
  $B^{\rm null}_\ell=\{0,0,0\}$ for the first column ($B_0$),
  $B^{\rm model}_\ell=\{B^{\rm bf}_0,B^{\rm bf}_2,0\}$ and
  $B^{\rm null}_\ell=\{B^{\rm bf}_0,0,0\}$ for the second column ($B_2$),
    $B^{\rm model}_\ell=\{B^{\rm bf}_0,0,B^{\rm bf}_4\}$ and
  $B^{\rm null}_\ell=\{B^{\rm bf}_0,0,0\}$ for the third column ($B_4$),
     $B^{\rm model}_\ell=\{B^{\rm bf}_0,B^{\rm bf}_2,B^{\rm bf}_4\}$ and
  $B^{\rm null}_\ell=\{B^{\rm bf}_0,0,0\}$ for the fourth column ($B_2+B_4$),
  where $B^{\rm bf}_{\ell=0,2,4}$ are best-fit theory curves. We note that the covariance matrix is highly correlated, thus the detection significance of $B_2+B_4$ pair is not equal to the sum of the individual $B_2$ and $B_4$ significances. Furthermore, we ignore the correlation between the bispectra and power spectra in our estimates, and consider only wavenumbers in the range $0.01<k/(\hMpc)<0.08$, yielding 62 triangle bins per multipole. We find a strong detection of the BOSS bispectrum monopole in all data chunks, and a somewhat less significant detection the higher multipoles in three out of four data chunks.
  }
\label{tab:snrt}
\end{table}

\begin{figure}
    \centering
    \includegraphics[width=\textwidth]{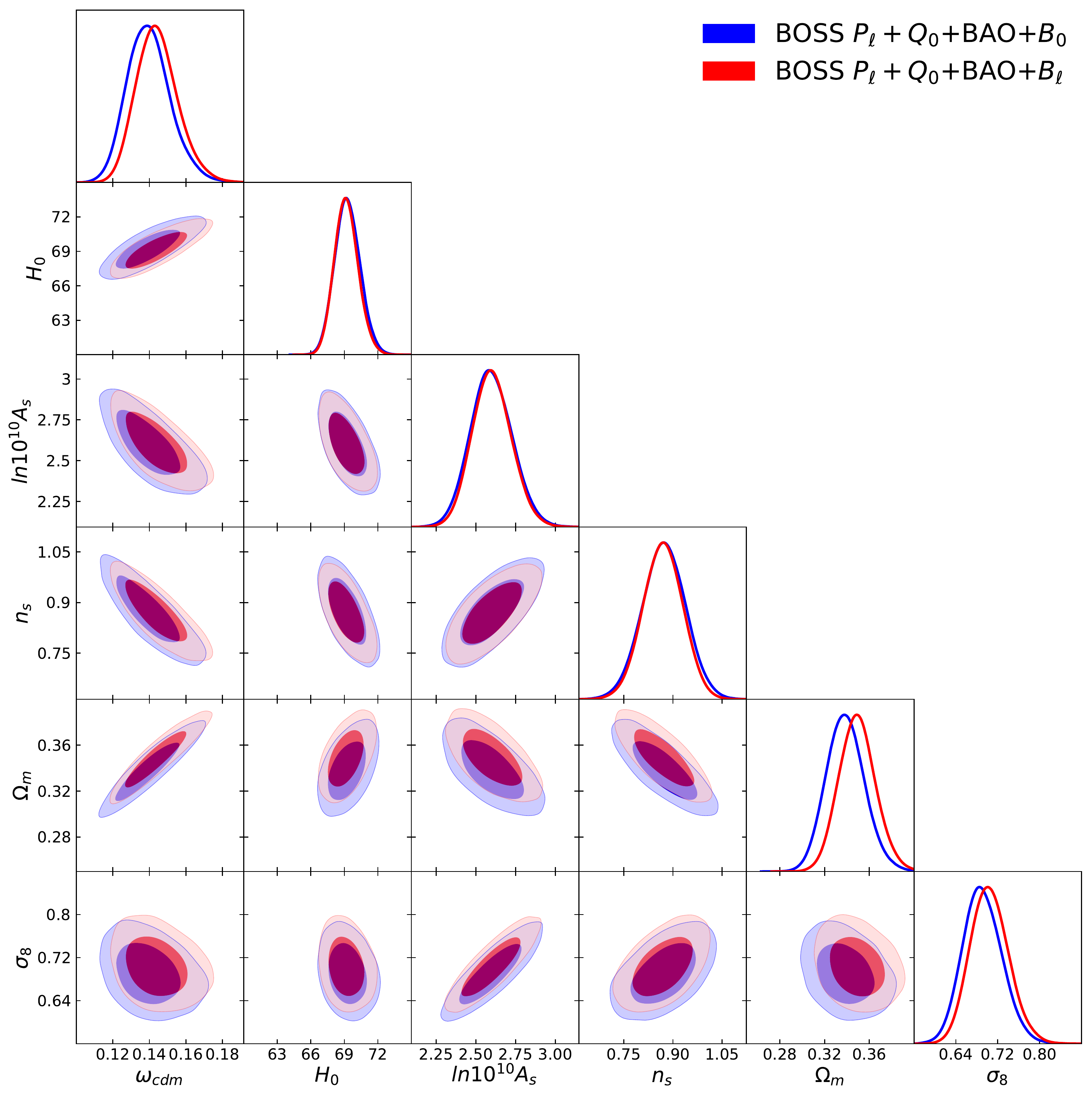}
    \caption{Constraints on $\Lambda$CDM cosmological parameters from the BOSS DR12 dataset. We compare results from the combined power spectrum, BAO, and bispectrum monopole ($\ell=0$) dataset (blue) and those adding the $\ell=2,4$ bispectrum multipoles (red). The inclusion of bispectrum multipoles is found to tighten parameter constraints only slightly, with most significant variation found in $n_s$ and $\Omega_m$.}
    \label{fig:boss_post_ns}
\end{figure}

\begin{figure}[h!]
    \centering
    \includegraphics[width=0.59\textwidth]{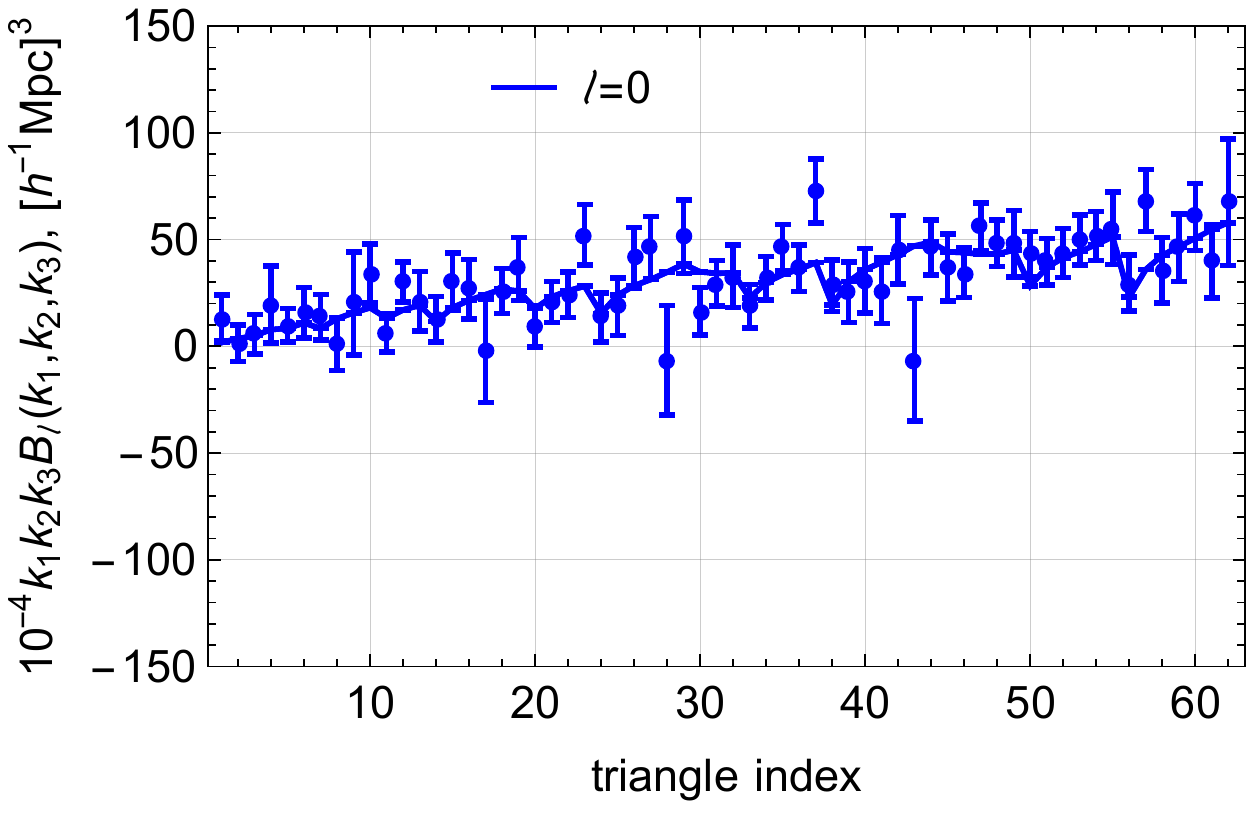}
    \includegraphics[width=0.59\textwidth]{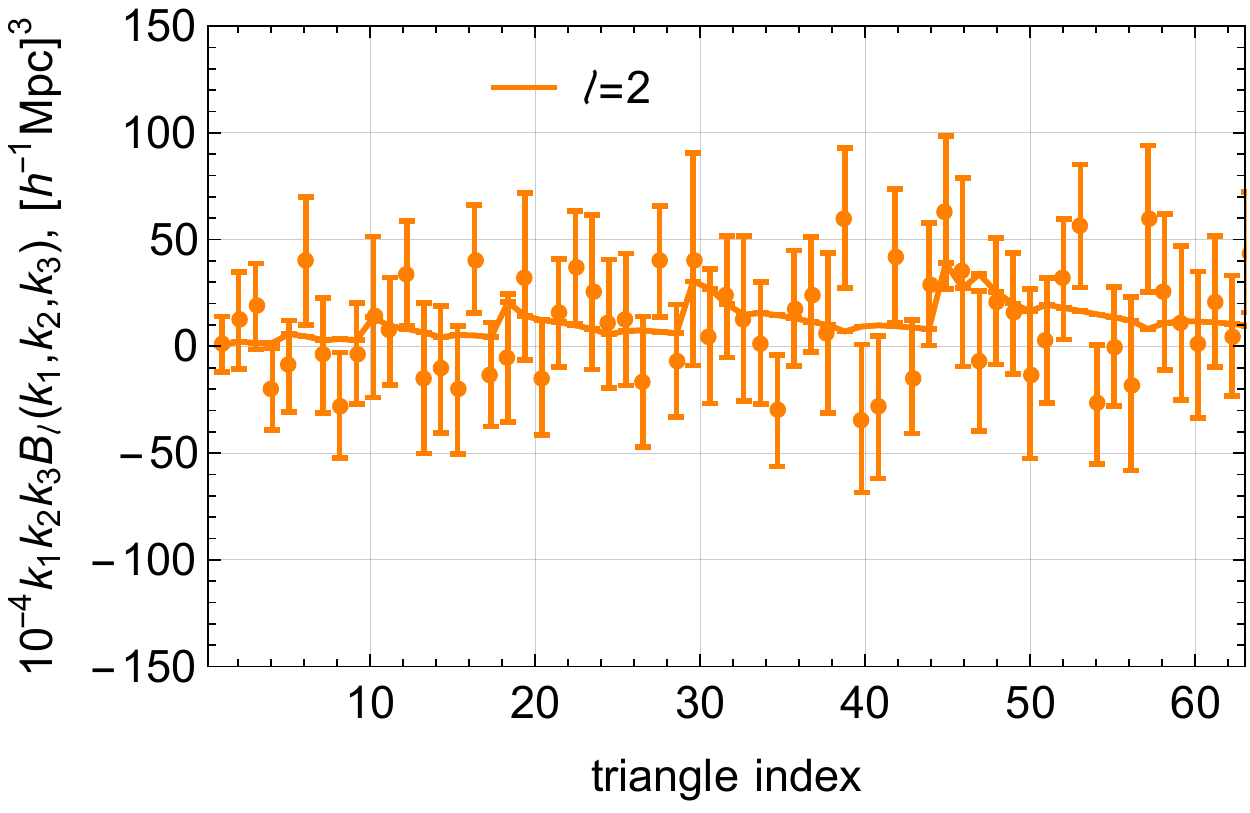}
    \includegraphics[width=0.59\textwidth]{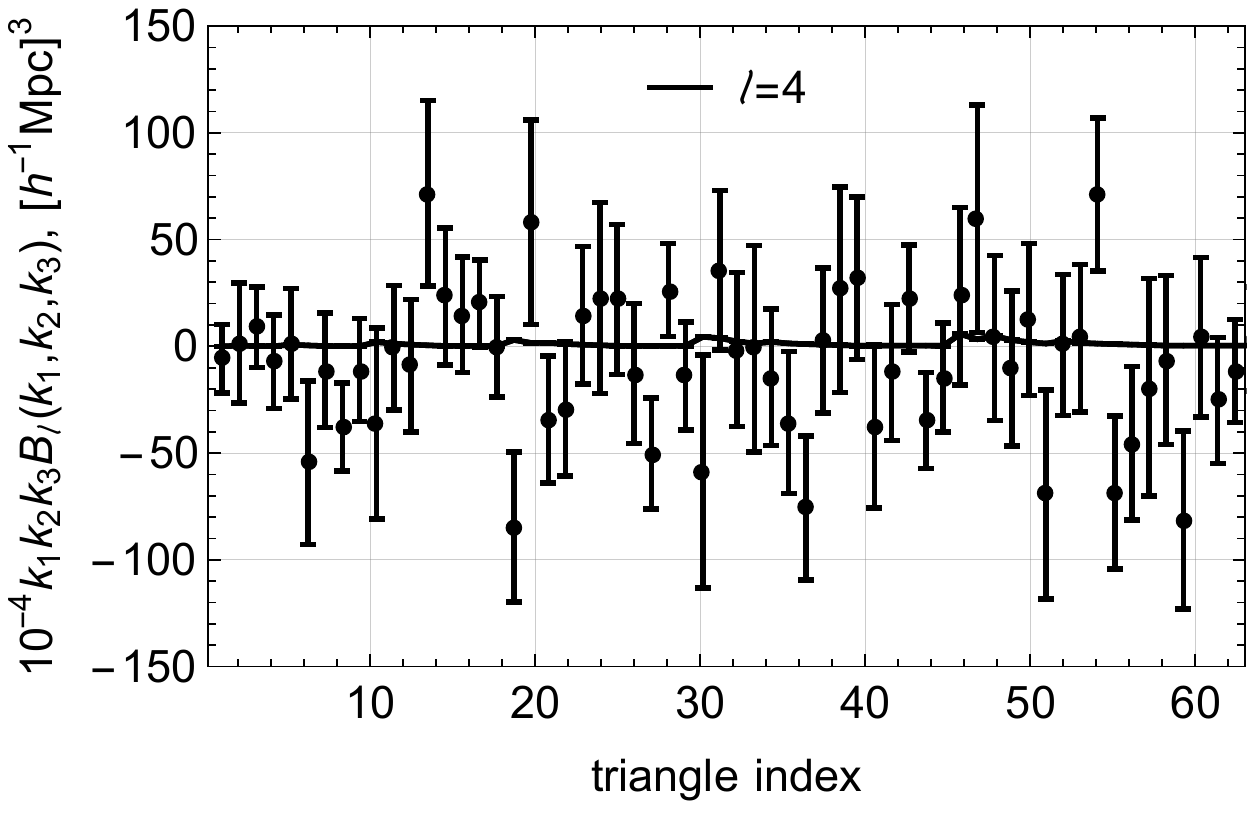}
    \caption{Comparison of the measured and theoretical galaxy bispectrum multipoles. We show the BOSS NGC high-z ($z=0.61$) data, along with the best-fit theory curves from our MCMC analysis. The top, middle, and bottom panels show the monopole, quadrupole, and hexadecapole respectively. Data are shown for $\kmax=0.08~\hMpc$ with all elements stacked (with smallest scales shown on the right). Errorbars correspond to diagonal elements of the covariance matrix, estimated from mocks. Though the signal of the higher-order BOSS multipoles is relatively small (see Tab.\,\ref{tab:snrt}), the model provides an excellent fit to the data, as evidenced by the simulation results in Fig.\,\ref{fig:ptc-plot}.}
    \label{fig:boss_bisp_data}
\end{figure}

We begin with a summary of our cosmological results. In this work, we have developed new window-free estimators for the bispectrum multipoles and applied them to the BOSS DR12 luminous red galaxy sample \cite{BOSS:2016wmc} (in two redshift bins and hemispheres), computing the monopole, quadrupole, and hexadecapole ($\ell=0,2,4$) of both the redshift-space power spectrum and bispectrum.  We additionally analyze the Alcock-Paczynski parameters from reconstructed power spectrum (following Ref.\,\citep{Philcox:2020vvt}), and the real-space power spectrum proxy $Q_0$~\cite{Ivanov:2021fbu} (see also Refs.\,\cite{Scoccimarro:2004tg,Hamilton:2000du,Tegmark:2003uf}). Our dataset matches that of our previous analysis~\cite{Philcox:2021kcw}, but supplemented 
with the bispectrum quadrupole and hexadecapole moments. For all the bispectrum moments used in this work, we focus on large-scale modes with $\kmax^B=0.08~\hMpc$, and limit ourselves with $\kmin^B=0.01~\hMpc$ to mitigate large-scale observation systematics. The power spectrum and bispectrum multipoles are measured with new maximum-likelihood estimators, as derived in \S\ref{sec:estim} (building on Refs.\,\cite{Philcox:2020vbm,Philcox:2021ukg}). These allow for robust comparison of theory and data without the need for window convolution.

In terms of theory, we use a tree-level perturbative model for the bispectrum multipoles (in the form introduced in Ref.~\cite{Ivanov:2021kcd}, and later used in Refs.\,\cite{Philcox:2021kcw,Cabass:2022wjy,Cabass:2022ymb}, {see also Refs.\,\citep{DAmico:2019fhj,DAmico:2022gki,DAmico:2022osl}}). 
We consistently fit the BOSS bispectrum multipole data, recomputing the theoretical templates for each set of cosmological parameters sampled in our MCMC chains. We focus on the minimal $\Lambda$CDM model and 
assume a BBN prior on the physical baryon density $\omega_b$~\cite{Aver:2015iza,Cooke:2017cwo,Ivanov:2019hqk}, with all other parameters fit directly from the BOSS data. Before analyzing the BOSS data, we test our fitting pipeline and estimators on a set of high-quality simulated galaxy catalogs, including the PT challenge mocks \citep{Nishimichi:2020tvu}. Our fits match these data well and we recover the true cosmological parameters in these cases, as shown in Fig.\,\ref{fig:ptc-plot} for the PT challenge data and the best-fit theory model. This implies that our pipeline for the bispectrum multipoles is adequate at the percent precision level, which even exceeds the statistical power of futuristic surveys.

Our main results are shown in Fig.~\ref{fig:boss_post_ns} and Tab.~\ref{tab:boss}. For comparison, we also display the constraints obtained from our previous BOSS likelihood that included only the bispectrum monopole \mbox{($\ell=0$)} moment \citep{Philcox:2021kcw}. The inclusion of the bispectrum multipole moments is found to have only a marginal effect on the cosmological parameter posteriors. Considering the $\Omega_m-\sigma_8$ plane, we find a slight reduction in the errorbars and a small posterior shift, which drives the clustering amplitude parameter $S_8\equiv\sigma_8 (\Omega_m/0.3)^{0.5}$ (at $z=0$) upwards by $\approx 0.6\sigma$. The largest effect can be seen in the marginalized $n_s$-posterior, which narrows by $\approx 10\%$ from the inclusion of $\ell=2,4$ galaxy bispectrum moments. All other one-dimensional posteriors on cosmological parameters typically shrink by $\lesssim 5\%$. These modest gains are a consequence of the relatively low signal-to-noise of the large-scale BOSS galaxy bispectrum multipoles. {As shown in Fig.~\ref{fig:boss_bisp_data} and in Tab.\,\ref{tab:snrt}, we could detect the higher order large-scale bispectrum multipoles only at $\approx 5\sigma$ in three out of the four BOSS data chunks. In comparison, the bispectrum monopole moment is detected typically at more than $10\sigma$ in all of the regions. This occurs due to the larger noise and reduced signal intrinsic to higher-order moments. We caution, however, that this $\Delta \chi^2$ detection metric does not fully reflect the impact on parameter constraints, for which one should use appropriate Fisher derivatives. We further note that we do not detect the higher order multipoles in the high-z SGC data chunk (which is small in volume), with the anisotropic clustering signal even being disfavored at around $2\sigma$. Whilst not significant, this result may be driven by neglecting the correlation with the power spectrum in our estimate, or by a statistical fluctuation.}

In addition, we remind that the particular one-dimensional parameter projections may not completely reflect changes in the full multi-dimensional posterior. In particular, the impact of the higher order multipole moments may be larger in extended cosmological models, analogous to the improvements found for the power spectrum \citep{Chudaykin:2020ghx}. The parameter improvements continue to be modest when we include a \textit{Planck} prior on the primordial power spectrum tilt $n_s$, as shown in the lower rows of Tab.~\ref{tab:boss}. Finally, it is worth stressing 
that the inclusion of the new data sets such as reconstructed power spectra, $Q_0$, and $B_\ell$ ($\ell=0,2,4$) yields significant improvements over the usual power spectrum-alone analysis.  Indeed, our final constraints on $\sigma_8$ are $\approx 30\%$ tighter than those from BOSS $P_\ell(k)$ alone, cf.~\cite{Philcox:2021kcw}.

\begin{figure}
    \centering
    \includegraphics[width=\textwidth]{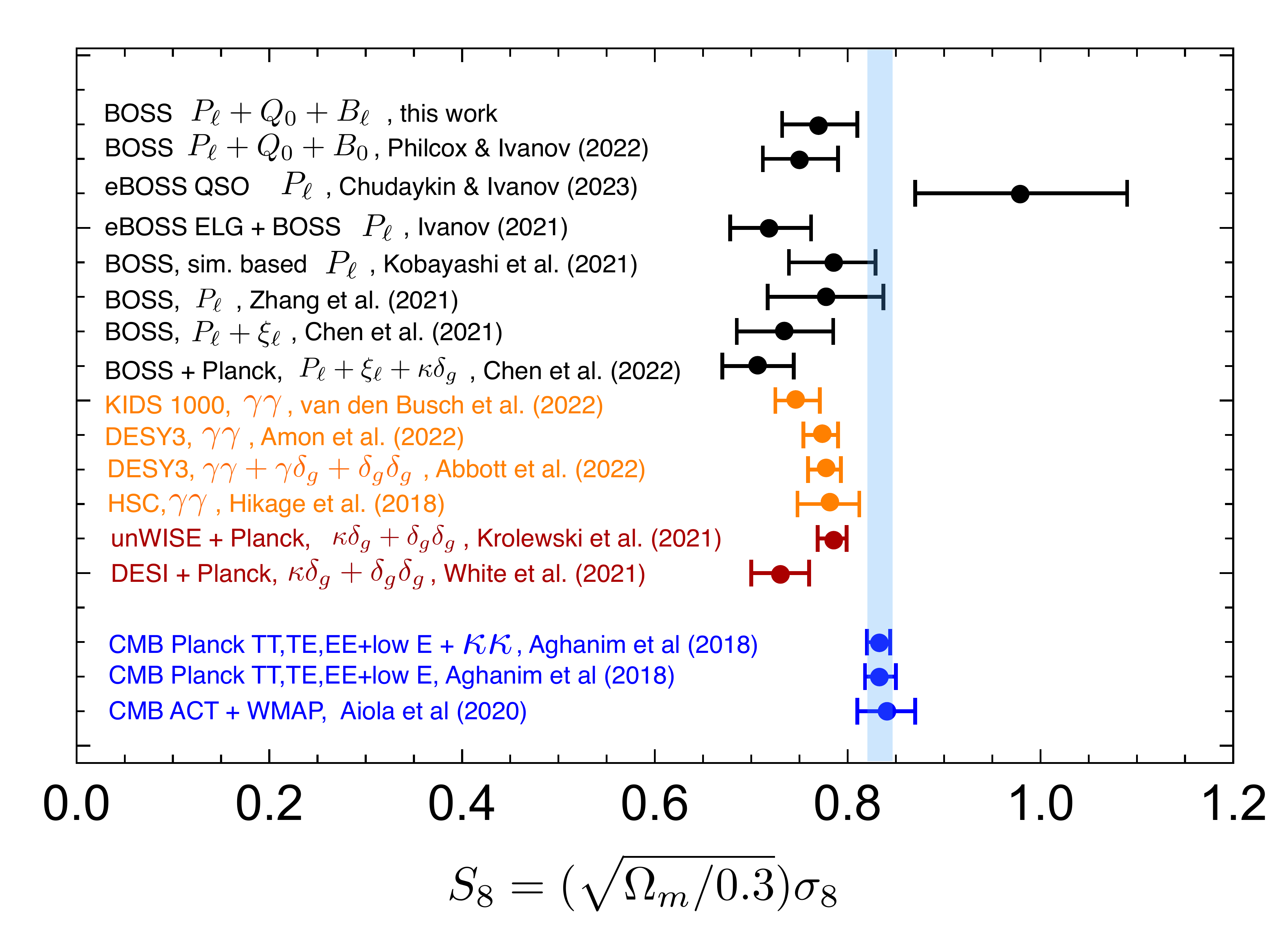}
    \caption{A compilation of some direct and indirect measurements of the growth parameter $S_8$, from spectroscopic surveys, weak lensing, and the CMB. Errorbars shown approximately correspond to the $68\%$ CL, and our measurement is shown in the top row. Further detail is given in Ref.~\cite{Chen:2022jzq} and the main text.}
    \label{fig:chart}
\end{figure}

To place our results in context, let us compare the optimal value of $S_8$ from our chains with those from other measurements. The direct measurements of this parameter from various weak lensing and galaxy clustering surveys (KIDS-1000~\cite{Busch:2022pcx}, DESY3~\cite{DES:2021bvc,DES:2021vln,DES:2021wwk}, HSC~\cite{HSC:2018mrq}, unWISE+\textit{Planck}~\cite{Krolewski:2021yqy}, DESI+\textit{Planck}~\cite{White:2021yvw}) are summarized in Fig.~\ref{fig:chart}. We particularly focus our attention on the full-shape anisotropic galaxy clustering probes in redshift space~\cite{Chen:2022jzq,Chen:2021wdi,Kobayashi:2021oud,Philcox:2021kcw,Ivanov:2021zmi,Chudaykin:2022nru}. For comparison, we also show there the prediction of the $\Lambda$CDM fit to the primary \textit{Planck}~\cite{Aghanim:2018eyx} and ACT+WMAP CMB~\cite{ACT:2020gnv} data, which may be considered an indirect probe of $S_8$. Our notation and choice of data sets follow those of Ref.~\cite{Chen:2022jzq}. Our measurement is fully consistent with those of other BOSS full-shape analyses, obtained both using perturbation  theory~\cite{Chen:2022jzq,Chen:2021wdi} and simulation-based frameworks~\cite{Kobayashi:2021oud}. We find a small (and relatively insignificant) tension between the $S_8$ measurements from ELG~\cite{Ivanov:2021zmi} and QSO samples~\cite{Chudaykin:2022nru} of the eBOSS survey~\cite{eBOSS:2020yzd}, which may be either due to residual systematics, or simply a statistical fluctuation. Finally, we point out that our $S_8$ posterior is broadly consistent with both CMB and various weak lensing probes. The latter two probes are in some $\sim 2\sigma$ disagreement with each other, which is often known as the $S_8$ tension (see Ref.~\cite{Abdalla:2022yfr} for a recent review).
We conclude that our measurement does not yield evidence for this tension.

\section{The Bispectrum Multipoles}\label{sec: bk-mult}
\subsection{Definition}
The galaxy bispectrum is defined as the three-point expectation of the overdensity, $\delta_g$: 
\beq
    \delD{\vk_{123}}B_{\rm ggg}(\vk_1,\vk_2,\vk_3) \equiv \av{\delta_g(\vk_1)\delta_g(\vk_2)\delta_g(\vk_3)}, 
\eeq
\citep[e.g.,][]{1980lssu.book.....P}, writing $\vk_{123}\equiv \vk_1+\vk_2+\vk_3$ for Dirac delta $\delta_{\rm D}$. In real-space, symmetry under translations and rotations forces the bispectrum to be a function only of three variables (usually chosen to be the side lengths $k_i\equiv |\vk_i|$); this implies $B_{\rm ggg}(\vk_1,\vk_2,\vk_3)\to B_{\rm ggg}(k_1,k_2,k_3)$. Redshift-space distortions break symmetry with respect to the line-of-sight $\hn$ (hereafter LoS), affording an additional two degrees of freedom to the bispectrum. Whilst this can be parametrize in a number of ways, a particularly well-motivated choice of variables are the angle of the triangle plane to the LoS, and the orientation of the triangle within the plane \citep[e.g.,][]{Verde:1998zr,Scoccimarro:2015bla,Colas:2019ret}.

In this approach, one can expand the bispectrum as a spherical harmonic series:
\beq
    B_{\rm ggg}(\vk_1,\vk_2,\vk_3) = \sum_{\ell=0}^\infty\sum_{m=-\ell}^\ell B_{\ell m}(k_1,k_2,k_3)Y_{\ell m}(\theta_{\vk},\phi_{\vk}),
\eeq
where $\theta_{\vk}$ and $\phi_{\vk}$ specify the aforementioned orientation. Though this basis is complete, measuring $B_{\ell m}$ is difficult, since the spherical harmonic cannot be separably decomposed into $\vk_1$, $\vk_2$, and $\vk_3$ pieces, yielding a non-factorizable estimator. This is not a problem for theoretical forecasts \citep[e.g.,][]{Mazumdar:2022ynd,Gualdi:2020ymf}, but severely limits application to observational data. Consequently, several works \citep[e.g.,][]{Scoccimarro:2015bla,Rizzo:2022lmh} have considered only the $m=0$ moment (independent of $\phi$), and set $\cos\theta\equiv\hk_3\cdot\hn$, additionally fixing $k_1\leq k_2\leq k_3$. This corresponds to representing the bispectrum as a Legendre series in $\theta$:
\beq\label{eq: Bk-Legendre}
    B_{\rm ggg}(\vk_1,\vk_2,\vk_3) \approx \sum_{\ell=0}^\infty B_{\ell}(k_1,k_2,k_3)\mathcal{L}_{\ell}(\hk_3\cdot\hn), \qquad (k_1\leq k_2\leq k_3)
\eeq
where $\mathcal{L}_\ell$ is a Legendre polynomial and $B_\ell$ the corresponding coefficient.\footnote{Some works \citep[e.g.,][]{Gualdi:2020ymf} have instead expanded the bispectrum as a \textit{double} Legendre series in the two angles. A separable choice would be to expand in, say, $\mathcal{L}_\ell(\hk_2\cdot\hn)$ and $\mathcal{L}_{\ell'}(\hk_3\cdot\hn)$; however, the corresponding coefficients are generally difficult to estimate robustly, since the two angles are not independent once the side-lengths are specified.} We note that \eqref{eq: Bk-Legendre} is not a strict equality, since the bispectrum contains higher-order moments (with $m\neq 0$) not captured within its formalism; in the below, we will instead define the multipoles directly as integrals over $\theta,\phi$.

\subsection{Idealized Estimators}
The decomposition of \eqref{eq: Bk-Legendre} can be used to construct estimators for the bispectrum multipoles, $B_\ell$. For an idealized periodic-box geometry (such as an $N$-body simulation), the conventional estimator for the bispectrum multipoles is given by 
\beq\label{eq: Bmult-scocc}
    \left.\widehat B^{abc}_\ell\right|_{\rm periodic} &\equiv& \frac{2\ell+1}{N_T^{abc}}\int_{\vk_1\vk_2\vk_3}\delD{\vk_{123}}\Theta^a(k_1)\Theta^b(k_2)\Theta^c(k_3)\\\nonumber
    &&\qquad\qquad\qquad\,\times\,\delta_g(\vk_1)\delta_g(\vk_2)\delta_g(\vk_3)\mathcal{L}_\ell(\hk_3\cdot\hn),
\eeq
where $\int_{\vk}\equiv (2\pi)^{-3}\int_{\vk}$ \citep{Scoccimarro:2015bla}. Here, $a\leq b\leq c$ specify a triplet of $k$-bins of finite radius, defined by $\Theta^i(k)$, which is unity if $k$ is in bin $i$, and zero else. \eqref{eq: Bmult-scocc} is simply an integral over three copies of the density field weighted by the Legendre polynomial in the longest side $\mathcal{L}_\ell(\hk_3\cdot\hn)$, with translation invariance enforced by the Dirac delta. This is normalized by the isotropic bin volume, defined by
\beq
    N_T^{abc} = \int_{\vk_1\vk_2\vk_3}\delD{\vk_{123}}\Theta^a(k_1)\Theta^b(k_2)\Theta^c(k_3).
\eeq
In this work, we regard \eqref{eq: Bmult-scocc} as the \textit{definition} of the binned bispectrum multipoles (rather than the approximate relation of \ref{eq: Bk-Legendre}). 

Theoretical predictions for the bispectrum multipoles can be similarly computed from the expectation of \eqref{eq: Bmult-scocc}:
\beq
    \left. B^{abc}_\ell\right|_{\rm theory} &\equiv&\frac{2\ell+1}{N_T^{abc}}\int_{\vk_1\vk_2\vk_3}\delD{\vk_{123}}\Theta^a(k_1)\Theta^b(k_2)\Theta^c(k_3)\\\nonumber
    &&\qquad\qquad\,\times\,B_{ggg}^{\rm theory}(\vk_1,\vk_2,\vk_3)\mathcal{L}_\ell(\hk_3\cdot\hn),
\eeq
for some theory model $B_{\rm theory}$ which is not yet averaged over angles. This will be discussed in \S\ref{sec:theory}.

In practice, we implement \eqref{eq: Bmult-scocc} by factorizing in $\vk_i$, following Ref.\,\citep{Scoccimarro:2015bla}. This is realized by rewriting the Dirac function as an exponential, yielding the asymmetric expression
\beq\label{eq: Bperiodic-estimator}
    \left.\widehat{B}^{abc}_\ell\right|_{\rm periodic} = \frac{2\ell+1}{N_T^{abc}}\int d\vx\,F_0^a(\vx)F_0^b(\vx)F_\ell^c(\vx), \quad N_T^{abc} = \int d\vx\,D^a(\vx)D^b(\vx)D^c(\vx), 
\eeq
using the definitions
\beq\label{eq: F-ell}
    F_\ell^i(\vx) \equiv \int_{\vk}e^{-i\vk\cdot\vx}\Theta^i(k)\delta(\vk)\mathcal{L}_\ell(\hk\cdot\hn), \qquad D^i(\vx) \equiv \int_{\vk}e^{-i\vk\cdot\vx}\Theta^i(k).
\eeq
Each piece can be straightforwardly evaluated using fast Fourier transforms (FFTs) with $N_g\log N_g$ complexity for $N_g$ grid points. If we had defined the redshift-space components using $Y_{\ell m}(\theta_{\vk},\phi_{\vk})$ rather than $\mathcal{L}_\ell(\hk_3\cdot\hn)$, (or some other choice) the expression would not factorize in the above manner, and computation would scale as $\mathcal{O}(N_g^3)$. 

In realistic surveys, the LoS is not fixed, but varies depending on which galaxies are being considered.\footnote{Strictly, a separate line-of-sight is required for each galaxy. The effects of assuming a single line-of-sight are small for typical survey sizes however \citep[cf.][]{Garcia:2020per,Philcox:2021tfv}.} In this case, we can adopt the `Yamamoto' prescription \citep{Yamamoto:2005dz,Scoccimarro:2015bla}, fixing $\hn$ to the direction vector of the galaxy associated to $\vk_3$. This corresponds to the replacement
\beq
    F_\ell^i(\vx) &\to& \int_{\vk}e^{-i\vk\cdot\vx}\Theta^i(k)\int d\vr\,e^{i\vk\cdot\vr}\delta(\vr)\mathcal{L}_\ell(\hk\cdot\hr) \\\nonumber
    &\equiv& \frac{4\pi}{2\ell+1}\sum_{m=-\ell}^{\ell}\int_{\vk}e^{-i\vk\cdot\vx}\Theta^i(k)Y_{\ell m}(\hk)\int d\vr\,e^{i\vk\cdot\vr}\delta(\vr)Y_{\ell m}^*(\hr),
\eeq
with the latter equality allowing for fast estimation using the spherical harmonic addition theorem.

\section{Window-Free Bispectrum Estimators}
\label{sec:estim}
\subsection{Motivation}
When applying the estimators described in \S\ref{sec: bk-mult} to observational data, we must specify the density field $\delta_g$. Usually, this is modelled by the pixelized field of ``data-minus-randoms''; $\delta_g(\vr) \propto n_g(\vr)-\alpha\,n_r(\vr)$, where $n_g$ is the observed galaxy density field and $n_r(\vr)$ is the random catalog (containing $1/\alpha$ times more particles than the galaxy catalog). Since both data and randoms are multiplied by the survey mask, conventional estimators will measure only the \textit{windowed} bispectrum, $B^{\rm win}_{ggg}$, rather than the true underlying statistic, $B_{ggg}$. Before bin integration, the two are related by the following convolution integral:
\beq\label{eq: Bk-conv}
    B_{\rm ggg}^{\rm win}(\vk_1,\vk_2,\vk_3) &=& \int_{\vp_1\vp_2\vp_3}\delD{\vp_{123}}\\\nonumber
    &&\qquad\qquad\,\times\,W(\vk_1-\vp_1)W(\vk_2-\vp_2)W(\vk_3-\vp_3)B_{\rm ggg}(\vp_1,\vp_2,\vp_3).
\eeq
To compare theory and data, we should similarly convolve the theory model. Due to its oscillatory nature, this is a difficult and time-consuming numerical operation (though see Ref.\,\citep{Pardede:2022udo} for a possible $\ell=0$ approach), thus the effect is often ignored or heavily simplified \citep[e.g.,][]{DAmico:2022osl,Gualdi:2018pyw,Gil-Marin:2014sta,Gil-Marin:2016wya,Gil-Marin:2014baa,DAmico:2019fhj,DAmico:2022gki}. This may lead to biases in data-analysis when large-scale modes (relevant to primordial non-Gaussianity studies) are included. 

A major goal of this work is the estimation of \textit{unwindowed} bispectrum multipoles. These are unbiased by the window function and can be robustly compared to theory models without the need to window-convolve the latter (via \ref{eq: Bk-conv}). Our approach follows Refs.\,\citep{Philcox:2020vbm,Philcox:2021ukg} for the power spectrum and $\ell=0$ monopole (as well as Ref.\,\citep{PhilcoxNpoint} for the higher-point CMB correlators), themselves inspired by early work on the subject in 
\citep{Tegmark:1997yq,Hamilton:1997kv,Hamilton:2005ma}.

\subsection{Binned Bispectrum Components}
To define unwindowed estimators, we must first express the true bispectrum $B_{ggg}(\vk_1,\vk_2,\vk_3)$ in terms of the quantity of interest: the set of bispectrum coefficients $b_{\alpha}\equiv B^{abc}_\ell$ (using $\alpha$ to denote the radial bin indices and multipole). This relation will then be used to form an estimator for $b_\alpha$ via maximum-likelihood methods. As an \textit{ansatz}, we will assume
\beq\label{eq: Bk-ansatz}
    B_{\rm ggg}(\vk_1,\vk_2,\vk_3) = \sum_\alpha \frac{b_\alpha}{\Delta_\alpha}\left[\Theta^a(k_1)\Theta^b(k_2)\Theta^c(k_3)\mathcal{L}_\ell(\hk_3\cdot\hn)+\text{5 perms.}\right].
\eeq
This is similar in form to the Legendre decomposition of \eqref{eq: Bk-Legendre}, but is defined for all arbitrary ordering of $\{\vk_1,\vk_2,\vk_3\}$, with the binning functions picking out the relevant permutation, such that we can represent the full bispectrum in terms of its binned components $b_\alpha$ with $a\leq b\leq c$. 
\eqref{eq: Bk-ansatz} includes a bin-specific normalization factor $\Delta_\alpha$; this takes a simple form for $\ell=0$ as in Ref.\,\citep{Philcox:2021ukg} but is more complex in general, as we show below, due to the omitted $\phi$ integrals and exchange symmetry.

Inserting \eqref{eq: Bk-ansatz} into the expectation of our idealized estimator \eqref{eq: Bmult-scocc} gives
\beq\label{eq: Bk-tmp}
    \av{\widehat B^{abc}_\ell} &=& \frac{2\ell+1}{N_T^{abc}}\int_{\vk_1\vk_2\vk_3}\delD{\vk_{123}}\Theta^a(k_1)\Theta^b(k_2)\Theta^c(k_3)\mathcal{L}_\ell(\hk_3\cdot\hn)\\\nonumber
    &&\,\times\,\sum_\beta \frac{b_\beta}{\Delta_\beta}\left[\Theta^{a'}(k_1)\Theta^{b'}(k_2)\Theta^{c'}(k_3)L_{\ell'}(\hk_3\cdot\hn)+\text{5 perms.}\right],
\eeq
where $\beta \equiv\{a',b',c',\ell'\}$. Assuming non-overlapping bins, the integral will be non-zero only when $\{a',b',c'\}$ is some permutation of $\{a,b,c\}$ (again restricting to $a'\leq b'\leq c'$). Invoking global rotational invariance, we can average over the LoS, making use of the relation:
\beq
    \int\frac{d\hn}{4\pi}\mathcal{L}_\ell(\hk_i\cdot\hn)\mathcal{L}_{\ell'}(\hk_j\cdot\hn) = \frac{\delta_{\rm K}^{\ell\ell'}}{2\ell+1}\mathcal{L}_\ell(\hk_i\cdot\hk_j).
\eeq
Writing out the permutations explicitly, this gives
\beq\label{eq: Bk-tmp2}
    \av{\hat B^{abc}_\ell} &=& \frac{1}{N_T^{abc}}\frac{b_\alpha}{\Delta_\alpha}\int_{\vk_1\vk_2\vk_3}\delD{\vk_{123}}\Theta^a(k_1)\Theta^b(k_2)\Theta^c(k_3)\\\nonumber
    &&\,\times\,\left\{\left[\mathcal{L}_\ell(\hk_1\cdot\hk_3)\left[\delta_{\rm K}^{bb'}\delta_{\rm K}^{ca'}+\delta_{\rm K}^{ba'}\delta_{\rm K}^{cb'}\right]\delta_{\rm K}^{ac'}+\mathcal{L}_\ell(\hk_2\cdot\hk_3)\left[\delta_{\rm K}^{aa'}\delta_{\rm K}^{cb'}\delta_{\rm K}^{ab'}\delta_{\rm K}^{ca'}\right]\delta_{\rm K}^{bc'}\right.\right.\\\nonumber
    &&\qquad\qquad\left.\left.+\,\delta_{\rm K}^{aa'}\delta_{\rm K}^{bb'}+\delta_{\rm K}^{ab'}\delta_{\rm K}^{ba'}\right]\delta_{\rm K}^{cc'}\right\}.
\eeq
The Kronecker deltas demarcate four scenarios: (1) $a\neq b\neq c$, (2) $a=b\neq c$, (3) $a\neq b=c$, (4) $a=b=c$. The latter two are more complex since they involve additional Legendre polynomials of two different $\vk$ vectors. To simplify these, we define the term:
\beq
    N_\ell^{abc} &\equiv& \int_{\vk_1\vk_2\vk_3}\delD{\vk_{123}}\Theta^a(k_1)\Theta^b(k_2)\Theta^c(k_3)\mathcal{L}_\ell(\hk_2\cdot\hk_3)\\\nonumber 
    &=&\frac{4\pi}{2\ell+1}\sum_{m=-\ell}^\ell\int d\vx\,\left[\int_{\vk_1}e^{-i\vk_1\cdot\vx}\Theta^a(k_1)\right]\left[\int_{\vk_2}e^{-i\vk_2\cdot\vx}\Theta^c(k_2)Y_{\ell m}(\hk_2)\right]\\\nonumber
    &&\qquad\,\times\,\left[\int_{\vk_3}e^{-i\vk_3\cdot\vx}\Theta^c(k_3)Y_{\ell m}^*(\hk_3)\right],
\eeq
rewriting the Dirac function as an exponential in the second line, allowing expression in terms of Fourier transforms. We note that $N_0^{abc}$ is just the isotropic bin volume $N_T^{abc}$. With the above definitions, we obtain the desired result $\av{\hat B^{abc}_\ell} = b_\alpha$ (\textit{i.e.}\ an unbiased estimator) subject to the following definition:
\beq
    \Delta_\alpha &\equiv& \begin{cases} 1 & a\neq b\neq c\\ 2 & a=b\neq c \\ \left(1+N_\ell^{abc}/N_T^{abc}\right) & a\neq b=c\\ 2\left(1+2N_\ell^{abc}/N_T^{abc}\right) & a=b=c.\end{cases}
\eeq
For $\ell=0$, this reduces to the symmetry factors used in \citep{Philcox:2021ukg} (1 for scalene, 2 for isosceles, 6 for equilateral). This calculation generalizes the standard bispectrum definition \eqref{eq: Bk-Legendre} to the binned bispectrum beyond the narrow bin limit (whence $a\neq b\neq c$ is guaranteed).

\subsection{Maximum-Likelihood Estimators}
We now consider the estimation of bispectrum coefficients $b_\alpha$, given their relation to the full bispectrum $B_{\rm ggg}(\vk_1,\vk_2,\vk_3)$. Following Refs.\,\citep{Philcox:2020vbm,Philcox:2021ukg,PhilcoxNpoint}, our pathway to this will be:
\begin{enumerate}
    \item Write down the likelihood for the observed pixellized data-minus-randoms field $\vec d$ in terms of the pixel correlators $\mathsf{C}_{ij}\equiv \av{d_id_j}$, $\mathsf{B}_{ijk}\equiv \av{d_id_jd_k}$, \textit{et cetera}, where $i,j,\cdots\in[1,N_{\rm pix}]$ are pixel indices.
    \item Express the relevant correlator (here $\B_{ijk}$) in terms of the coefficients of interest, \textit{i.e.}\ the binned bispectrum multipoles $b_\alpha$.
    \item Maximize the log-likelihood with respect to $b_\alpha$ forming a quasi-optimal estimator.
    \item Simplify the resulting form such that it can be efficiently implemented on data using FFTs.
\end{enumerate}

In the weakly non-Gaussian regime, the likelihood of the data is given by the Edgeworth expansion \citep[e.g.,][]{Sellentin:2017aii}
\beq\label{eq: edgeworth}
    -\log L[\vec d] = -\log L_G[\vec d] -  \frac{1}{3!}\B^{ijk}\left(h_ih_jh_k-h_i\mathsf{C}^{-1}_{jk}-h_j\mathsf{C}^{-1}_{ik}-h_k\mathsf{C}^{-1}_{ij}\right)+\cdots
\eeq
where $L_G$ is the Gaussian piece (which we do not need here), and $h_i\equiv \mathsf{C}^{-1}_{ij}d^j$ is the Wiener-filtered data. In this formalism, the optimal estimator for $b_\alpha$ (which enters linearly in $\mathsf{B}^{ijk}$) is given by
\beq
    \widehat{b}_\alpha = \sum_{\beta}\left(F^{-1}\right)_{\alpha\beta}\widehat{b}_\beta^{\rm num},
\eeq
defining the numerator and normalization:
\beq\label{eq: ML-estimator}
    \widehat{b}_\alpha^{\rm num} &=& \frac{1}{6}\frac{\partial\mathsf{B}^{ijk}}{\partial b_\alpha}\left[h_ih_jh_k-\left(h_i\mathsf{C}^{-1}_{jk}+\text{2 perms.}\right)\right]\\\nonumber
    F_{\alpha\beta} &=& \frac{1}{6}\frac{\partial\mathsf{B}^{ijk}}{\partial b_\alpha}\mathsf{C}^{-1}_{il}\mathsf{C}^{-1}_{jm}\mathsf{C}^{-1}_{kn}\frac{\partial\mathsf{B}^{lmn}}{\partial b_\beta}.
\eeq
This is just the maximum likelihood solution of \eqref{eq: edgeworth}.

In our case, the three-point function can be written as a Fourier-transform of the full redshift-space bispectrum $B_{ggg}(\vk_1,\vk_2,\vk_3)$, noting that $d_i\equiv n(\vr_i)\delta_g(\vr_i)$ for background density $n(\vr)$:
\beq
    \B^{ijk} = n(\vr_i)n(\vr_j)n(\vr_k)\int_{\vk_1\vk_2\vk_3}e^{i\vk_1\cdot\vr_i+i\vk_2\cdot\vr_j+i\vk_3\cdot\vr_k}\delD{\vk_{123}}B_{ggg}(\vk_1,\vk_2,\vk_3),
\eeq
Inserting \eqref{eq: Bk-ansatz}, we can write the cumulant derivative as
\beq\label{eq: cum-deriv}
    \frac{\partial \B^{ijk}}{\partial b_\alpha} &=& \frac{n(\vr_i)n(\vr_j)n(\vr_k)}{\Delta_\alpha}\int_{\vk_1\vk_2\vk_3}\left[\Theta^a(k_1)\Theta^b(k_2)\Theta^c(k_3)\mathcal{L}_\ell(\hk_3\cdot\hn)+\text{5 perms.}\right]\\\nonumber
    &&\qquad\qquad\qquad\qquad\,\times\,e^{i\vk_1\cdot\vr_i+i\vk_2\cdot\vr_j+i\vk_3\cdot\vr_k}\delD{\vk_{123}}.
\eeq
Under the Yamamoto approximation, we fix the LoS to be $\hn=\hr_3$, as above.

Inserting the above results into \eqref{eq: ML-estimator}, the numerator of the bispectrum estimator is found to be:
\beq\label{eq: q-alpha-binned}
    \widehat{b}^{\rm num}_\alpha &=& \frac{1}{\Delta_\alpha}\int d\vr\,\bigg[g^{a}_0[\vd](\vr)g^{b}_0[\vd](\vr)g^c_\ell[\vd](\vr)-\left(g^a_0[\vd](\vr)\av{g^b_0[\va](\vr)\tilde g^c_\ell[\va](\vr)}+\text{2 perms.}\right)\bigg],
\eeq
subject to the definitions
\beq\label{eq: g-alpha-binned}
    g^a_\ell[\vy](\vr) = \int_{\vk}e^{-i\vk\cdot\vr}\Theta^a(k)\int d\vr' e^{i\vk\cdot\vr'}n(\vr')[\Hi\vy](\vr')\mathcal{L}_\ell(\hk\cdot\hr')\\\nonumber
    \tilde g^a_\ell[\vy](\vr) = \int_{\vk}e^{-i\vk\cdot\vr}\Theta^a(k)\int d\vr' e^{i\vk\cdot\vr'}n(\vr')[\Ai\vy](\vr')\mathcal{L}_\ell(\hk\cdot\hr').
\eeq
$g_0^a$ is equal to the $g^a$ function of Ref.\,\citep{Philcox:2021ukg}. This is closely linked to the $F_\ell$ functions found in the ideal estimator \eqref{eq: F-ell}, but now includes the survey mask and custom weighting functions. Two points are of note: (a) we replace the $\mathsf{C}^{-1}$ Wiener filtering by a more general weighting $\mathsf{H}^{-1}$; (b) we introduce a set of random maps $\vec a$ with known covariance $\mathsf{A}$ following Ref.\,\citep{Smith:2006ud}. The former allows for a simple-to-implement estimator (since the full pixel covariance is difficult to compute and harder still to invert), and the latter allows one to compute the one-point terms via Monte Carlo summation (removing the need for a direct sum which has a prohibitive $\mathcal{O}(N_{\rm pix}^2)$ scaling).

Exploiting spherical harmonic factorizations, the two terms in \eqref{eq: g-alpha-binned} can be written in terms of forward and reverse Fourier-transforms $\mathcal{F}$ and $\mathcal{F}^{-1}$:
\beq
    g_\ell^a[\vy](\vr) &\equiv& \frac{4\pi}{2\ell+1}\sum_{m=-\ell}^\ell\ift{\Theta^a(k)Y^*_{\ell m}(\hk)\ft{n\Hi\vy\,Y_{\ell m}}(\vk)}(\vr)\\\nonumber
    \tilde g_\ell^a[\vy](\vr) &\equiv& \frac{4\pi}{2\ell+1}\sum_{m=-\ell}^\ell \ift{\Theta^a(k)Y^*_{\ell m}(\hk)\ft{n\Ai\vy\,Y_{\ell m}}(\vk)}(\vr).
\eeq

The second part of the estimator is a data-independent normalization (or Fisher) matrix, $F_{\alpha\beta}$. This acts to remove correlations between bins and multipoles and can be efficiently estimated via Monte Carlo methods. In the limit of ideal weighting ($\mathsf{H}^{-1}\to \mathsf{C}^{-1}$) and vanishing non-Gaussianity, the bispectrum covariance is equal to $F^{-1}$. As in \citep{Philcox:2021ukg}, this takes the form
\beq\label{eq: fish-applied}
    F_{\alpha\beta} = \frac{1}{12}\left(\av{\phi_{\alpha}^i\Hi_{il}\tilde\phi_{\beta}^l}-\av{\phi_{\alpha}^i}\Hi_{il}\av{\tilde\phi_\beta^l}\right),
\eeq
with $\phi_{\alpha}^i[\va] = \mathsf{B}^{ijk}_{,\alpha}\Hi_{jj'}\Hi_{kk'}a^{j'}a^{k'}$ and analogously for $\tilde\phi$ with $\Hi\to\Ai$. \eqref{eq: fish-applied} can be implemented by applying the linear map $\mathsf{H}^{-1}$ to $\tilde\phi$ then summing the result (multiplied by $\phi$) in pixel-space. Once again, the expectations can be computed by summation over Monte Carlo realizations $\vec a$ with known covariance $\mathsf{A}$ (e.g., Gaussian random fields).

With the above form for the cumulant derivative \eqref{eq: cum-deriv}, we can write the $\phi$ field explicitly in terms of Fourier transforms:
\beq\label{eq: phi-applied}
    \phi_\alpha^i[\va] &=& \B_{,\alpha}^{ijk}\Hi_{jj'}\Hi_{kk'}a^{j'}a^{k'}\\\nonumber
    &=& \frac{n(\vr_i)}{\Delta_\alpha}\int d\vr\,\int_{\vk_1\vk_2\vk_3}e^{i\vk_1\cdot\vr_i}\left[\Theta^a(k_1)\Theta^b(k_2)\Theta^c(k_3)\mathcal{L}_\ell(\hk_3\cdot\hn)+\text{5\,perms.}\right]\\\nonumber
    &&\qquad\qquad\,\times\,e^{-i(\vk_{123})\cdot\vr}[n\Hi\va](\vk_2)[n\Hi\va](\vk_3)\\\nonumber
    &=&\frac{2n(\vr_i)}{\Delta_\alpha}\left\{\ift{\Theta^a(k)\ft{g_0^b[\va]g_\ell^c[\va]}(\vk)}(\vr_i)+(a\leftrightarrow b)\right.\\\nonumber
    &&\qquad\qquad\left.\,+\,\ift{\Theta^c(k)\mathcal{L}_\ell(\hk\cdot\hr_i)\ft{g_0^a[\va]g_0^b[\va]}(\vk)}(\vr_i)\right\},
\eeq
with an analogous form for $\tilde\phi_\alpha$ involving $\tilde g_\ell^n$. The final term involves a Legendre polynomial; using spherical harmonic decompositions, this can be simplified to yield the form:
\beq
    \left.\phi_\alpha^i[\va]\right|_{\rm III} = \frac{2n(\vr_i)}{\Delta_\alpha}\frac{4\pi}{2\ell+1}\sum_{m=-\ell}^\ell Y_{\ell m}(\hr_i)\ift{\Theta^c(k)Y_{\ell m}^*(\hk)\ft{g_0^a[\va]g_0^b[\va]}(\vk)}(\vr_i).
\eeq

Collecting results, the full estimator for the bispectrum is given by
\beq\label{eq: Bk-mult-unwindowed}
    \widehat b_\alpha = \sum_{\beta}F_{\alpha\beta}^{-1}\widehat b^{\rm num}_\beta.
\eeq
This is unbiased for any choice of $\mathsf{H}^{-1}$, unwindowed, and, for $\mathsf{H}^{-1}\approx \mathsf{C}^{-1}$, close-to optimal (partly due to the inclusion of a linear term \citep[cf.][]{Smith:2006ud}). These properties are derived formally in \citep{Philcox:2021ukg}. Both the numerator and Fisher matrix can be efficiently computed using $N_{\rm mc}$ Monte Carlo simulations, with the finite number of simulations incurring an error proportional to $\sqrt{1+1/N_{\rm mc}}$. Whilst the latter is computationally expensive (requiring $\mathcal{O}(N_{\rm bins})$ Fourier transforms), it only has to be estimated once for a given survey geometry. We will discuss the specifics of our implementation in \S\ref{sec:likelihood}. A public \textsc{Python} implementation can be found online.\footnote{\href{https://github.com/oliverphilcox/Spectra-Without-Windows}{GitHub.com/OliverPhilcox/Spectra-Without-Windows}.}

\section{Theory Model Overview}
\label{sec:theory}

\subsection{Idealized Form}
To model the galaxy bispectrum models, we will use the tree-level theory introduced in Ref.~\cite{Ivanov:2021kcd} (see also~\cite{Scoccimarro:2000sn,Sefusatti:2006pa,2010MNRAS.406.1014S,Baldauf:2014qfa,Desjacques:2018pfv,Oddo:2019run,Oddo:2021iwq,Eggemeier:2018qae,Eggemeier:2020umu,Eggemeier:2021cam}). At a redshift $z$, the redshift-space bispectrum is the sum of three contributions:
\be
B_{\rm ggg}(\k_1,\k_2,\k_3)= B_{211}(\k_1,\k_2,\k_3)+B_{\rm ctr}(\k_1,\k_2,\k_3)
+B_{\rm stoch}(\k_1,\k_2,\k_3)\,,
\ee
where $B_{211}$ is the standard deterministic mode-coupling contribution,
\be
\begin{split}
B_{211}(\k_1,\k_2,\k_3) = 2Z_1(\k_1)Z_1(\k_2)Z_2(\k_1,\k_2)P_{11}(k_1)P_{11}(k_2)+\text{2 cyc.}\,.
\end{split} 
\ee
Here $P_{11}(k,z)$ is the linear matter power spectrum at redshift $z$, and the redshift-space
perturbation theory kernels are given by \citep[cf.,][]{Bernardeau:2001qr}
\bseq 
\begin{align}
&Z_1(\k)  = b_1+f\mu^2\,,\\
&Z_2(\k_1,\k_2)  =\frac{b_2}{2}+b_{\mathcal{G}_2}\left(\frac{(\k_1\cdot \k_2)^2}{k_1^2k_2^2}-1\right)
+b_1 F_2(\k_1,\k_2)+f\mu^2 G_2(\k_1,\k_2)\notag\\
&\qquad\qquad\quad~~+\frac{f\mu k}{2}\left(\frac{\mu_1}{k_1}(b_1+f\mu_2^2)+
\frac{\mu_2}{k_2}(b_1+f\mu_1^2)
\right)
\,,\\
& F_2(\k_1,\k_2)=\frac{5}{7}
+\frac{1}{2}\left(
\frac{(\k_1\cdot \k_2)}{k_1^2}
+\frac{(\k_1\cdot \k_2)}{k_2^2}
\right)+\frac{2}{7}\frac{(\k_1\cdot \k_2)^2}{k_1^2 k_2^2}\,,\\
& G_2(\k_1,\k_2)=\frac{3}{7}
+\frac{1}{2}\left(
\frac{(\k_1\cdot \k_2)}{k_1^2}
+\frac{(\k_1\cdot \k_2)}{k_2^2}
\right)+\frac{4}{7}\frac{(\k_1\cdot \k_2)^2}{k_1^2 k_2^2}\,,
\end{align} 
\eseq
where we introduced the following angles with respect to the line of sight directions: $\mu_i\equiv (\k_i\cdot \hat{\z})/k_i$ and $\mu\equiv (\k\cdot \hat{\z})/k$, $\k\equiv\k_1+\k_2$. Additionally, $f$ is the logarithmic growth factor, 
\be
f=\frac{d\ln D_+}{d\ln a}\,,
\ee
where $D_+$ denotes the usual growth rate and $a$ is the scale factor of the Friedmann-Lemaitre-Robertson-Walker metric. The free coefficients $b_1$, $b_2$, and $b_{\mathcal{G}_2}$ capture linear, quadratic, and tidal bias between galaxies and matter~\cite{Scoccimarro:1996jy,Bernardeau:2001qr,McDonald:2009dh,Mirbabayi:2014zca,Assassi:2014fva,Senatore:2014eva,Desjacques:2016bnm}. 

The second ingredient of our model is the counterterm contribution which is, essentially, a phenomenological term meant to capture the large-scale limit of non-linear redshift space distortions (``fingers-of-God'' (FoG)~\cite{Jackson:2008yv}). In the EFTofLSS, the higher derivative counterterms capture the backreaction effect induced by short-scale non-linearities. In the presence of RSD, this effect is dominated by stochastic virial velocities, which make up FoG. The physical distance scale associated with these velocities, $\sigma_v\sim 5$~[$h^{-1}$Mpc], is parametrically larger than the other scales in the EFT expansion, hence the RSD counterterms are important even on very large scales where 
the usual one-loop EFT corrections due to mode coupling are suppressed. For this reason we take the FoG counterterms into account but neglect the other one-loop corrections, effectively using a higher-order Taylor expansion for $\sigma_v$. In practice, our counterterm model amounts to modifying the kernel $Z_1$ as
\be
Z_1\to  Z^{\rm FoG}_1 = Z_1+\delta Z_1 = b_1 + f\mu^2 -c_1\mu^2 \left(\frac{k}{k^r_{\rm NL}}\right)^2\,,
\ee 
where $k_{\rm NL}^r=0.3~\hMpc$ 
is the RSD cutoff for the Red Luminous Galaxies~\cite{Nishimichi:2020tvu,Ivanov:2021zmi}.
We have found that this phenomenological model is sufficient to capture the leading effect of FoG on large scales. As one moves to shorter scales, a full set of counterterms becomes necessary, along with the appropriate one-loop corrections, as demonstrated in Ref.~\cite{Philcox:2022frc}. 

The third piece of our model is the stochastic contribution 
\be
B_{\rm stoch}(\k_1,\k_2,\k_3)= Z_1(\k_1)\frac{P_{11}(k_1)}{\bar n}\left(b_1 B_{\rm shot} +
f\mu^2 (1+P_{\rm shot})\right)+\frac{1+A_{\rm shot}}{\bar n^2}\,,
\ee
where $\bar n$ is the galaxy number density, and $A_{\rm shot}, B_{\rm shot}, P_{\rm shot}$ are free $\mathcal{O}(1)$ shot-noise parameters that capture deviations from Poissonian stochasticity. Note that mathematical consistency requires that the $P_{\rm shot}$ parameter is the same as that appearing in the power spectrum model. We additionally note that, in contrast to~\cite{Ivanov:2021kcd}, we do not make any assumptions on $A_{\rm shot}$, and keep this parameter free in the fit.

The last purely theoretical ingredient of our model is infrared (IR) resummation, which captures the non-linear evolution of baryon acoustic
oscillations~\cite{Crocce:2007dt,Senatore:2014via,Baldauf:2015xfa}. This is implemented using the prescription outlined in Refs.~\cite{Ivanov:2021kcd,Ivanov:2018gjr,Blas:2016sfa,Vasudevan:2019ewf}, developed within the context of time-sliced perturbation theory~\cite{Blas:2015qsi}.

\subsection{Observational Effects}
Two practical effects must also be taken into account in our model. The first is the coordinate distortion imprinted by the assumption of a fiducial cosmology (known as the Alcock-Paczynski effect, when applied to the shifts of the BAO peak ~\cite{Alcock:1979mp}). The relationship between the true underlying wavenumbers and angles ($q,\nu$) and the observed wavenumbers and angles $(k,\mu)$ is given by
\be
\begin{split}\label{eq:AP_k_mu}
	q^2&=k^2\left[
	\a_\parallel^{-2}
	\mu^2+
	\a_\perp^{-2}
	(1-\mu^2)\right]\,,\\
	\nu^2&=
	\a_\parallel^{-2}
	\mu^2\left[
	\a_\parallel^{-2}
	\mu^2+
	\a_\perp^{-2}
	(1-\mu^2)\right]^{-1}\,,
\end{split}
\ee
where 
\be
\alpha_\parallel = \frac{H_{\rm fid}(z)}{H_{\rm true}(z)}
\frac{H_{0,\rm true}}{H_{0,\rm fid}}
\,,\quad \alpha_\perp = \frac{D_{\rm true, A}(z)}{D_{\rm fid, A}(z)}
\frac{H_{0,\rm true}}{H_{0,\rm fid}}\,,
\ee
for angular diameter distance $D_{\rm A}$ and Hubble parameter $H$. Note that we have explicitly taken into account that wavenumbers are measured in units of $h\,\mathrm{Mpc}^{-1}$, yielding additional factors $H_{0,\rm true}/H_{0,\rm fid}$.
The bispectrum multipoles in physical redshift space are then given by~\cite{Song:2015gca} (see \S\ref{sec: bk-mult})
\be
\label{eq:bimultidef}
\begin{split}
&B_{\ell}(k_1,k_2,k_3)\\
&=\frac{2\ell+1}{2\alpha^2_\parallel\alpha^4_\perp}
\int_{0}^{2\pi} \frac{d\phi}{2\pi} \int_{-1}^{1}d\mu_3~ 
\mathcal{L}_{\ell}(\mu_3)~B_{\rm ggg}(q_1[k_1,\mu_1],q_2[k_2,\mu_2],q_3[k_3,\mu_3],\nu_1[\mu_1],\nu_2[\mu_2],\nu_3[\mu_3])
\,,
\end{split}
\ee
where $\mu_1,\mu_2$ are defined by $\mu_3$ and $\phi$. The observed angles being subject to~\eqref{eq:AP_k_mu}. In what follows we will focus on the $\ell=0,2,4$ moments. Higher order moments are also present, but they generate negligible signal on large scales, and can thus be ignored for the purposes of this paper.

The last observational effect is related to the discrete sampling of Fourier modes. We account for this effect following Ref.\,\cite{Ivanov:2021kcd} (with alternative binning methods discussed in Refs.~\cite{Oddo:2019run,Oddo:2021iwq,Eggemeier:2021cam,Rizzo:2022lmh}). Our method consists of two steps. As a first step (known as the ``continuum approximation''), 
one assumes that there is an infinitely dense continuum of Fourier modes, in which case the binning effects simplify to an integration of the bispectrum model over the chosen wavenumber bins. As a second step, deviations from the continuum approximation are taken into account by means of ``discreteness weights'',  defined as the ratio between the true binned bispectrum built out of discrete Fourier modes, and its continuous approximation, \textit{i.e.}\ 
\be
w=\frac{\hat B_{\ell,\rm disc}}{\hat B_{\ell,\rm int}}\,, 
\ee
where $\hat  B_{\ell,\rm int}$ is the bin-integrated bispectrum,
and $\hat  B_{\ell,\rm disc}$ is the explicitly-computed bispectrum model calculated on a discrete $k$-grid. Note that the angular integral~\eqref{eq:bimultidef} is replaced with a discrete sum over the available angular modes in this case.
The discreteness weights $w$ (which are expensive to compute) are defined for some fiducial cosmology that is consistent with the data. The residual cosmology-dependence of the weights 
is quite weak, and in principle, can be taken into account 
iteratively~\cite{Ivanov:2021kcd}. All in all, our theory model is given by 
\be 
B^{\rm th}_\ell = w_\ell(k_1,k_2,k_3)B_\ell^{\rm int}(k_1,k_2,k_3)\,.
\ee

\section{Data and Likelihood}
\label{sec:likelihood}

This paper uses three different types of data and corresponding likelihoods. First, we will analyze mock galaxy clustering data from the PT Challenge and Nseries mocks, with the former boasting huge volume and the latter including BOSS observational effects. In the second part of the paper, we analyze the observed BOSS DR12 LRG clustering data. 

\subsection{PT Challenge}
The PT Challenge simulation suite was created to test analytic modeling of the large-scale clustering of BOSS-like galaxies at the per-mile level~\cite{Nishimichi:2020tvu}, covering a cumulative volume of $566$ ($h^{-1}$Gpc)$^3$. These are periodic box simulations that are free of many observational effects, such as those of the lightcone (radial selection), window function, and fiber collisions. The mocks, however, include the Alcock-Paczynski effect. The publicly available 
simulation suite consists of 10 independent realizations with three snapshots at $z=0.38,0.51,0.61$. In this work, we will 
focus on a single snapshot at $z=0.61$, which matches the properties of the ``high-z'' BOSS DR12 data chunk. This dataset has been used to validate various analyses of EFT-based theoretical models for the galaxy power spectra and bispectra in Refs.~\cite{Nishimichi:2020tvu,Chudaykin:2020ghx,Ivanov:2021kcd,Ivanov:2021fbu,Philcox:2022frc,Chen:2020fxs}. 
Here, we extend these analyses to the galaxy bispectrum 
multipole moments. Our full data vector is given by
\be
\label{eq:datav}
\{P_0,P_2,P_4,Q_0,B_0,B_2,B_4\}\,, 
\ee
where $P_{\ell}$ ($\ell=0,2,4$) are the galaxy power spectrum multipoles with $\kmax^P=0.16~\hMpc$, $Q_0\equiv P_0-\frac{1}{2}P_2+\frac{3}{8}P_4$ is the real space galaxy power spectrum proxy (taken for $k_{\rm min}^Q=0.16~\hMpc$ and $\kmax^Q=0.4~\hMpc$), and $B_\ell$ ($\ell=0,2,4$) are the bispectrum multipole moments taken for $k_{\rm min}^B=0.01~\hMpc$ and $k_{\rm max}^B = 0.08~\hMpc$, and estimated using the periodic-box estimators of \eqref{eq: Bmult-scocc}. 

The power spectrum likelihood for $P_\ell$ and $Q_0$ has been discussed in detail in~\cite{Ivanov:2021fbu},  with that of the tree-level bispectrum monopole considered in~\cite{Ivanov:2021kcd}. Note that these scale cuts have been chosen by requiring the parameter estimation from PT Challenge mocks to be unbiased. In principle, one could measure the scale cut $\kmax$ without knowing the true underlying cosmology, e.g., using the theoretical error approach~\cite{Baldauf:2016sjb,Chudaykin:2020hbf}. 

In this work, we assume a Gaussian likelihood for the data vector~\eqref{eq:datav} with the covariance matrix computed in the Gaussian tree-level approximation, as verified for the power spectrum and the tree-level bispectrum likelihood in Ref.~\cite{Ivanov:2021kcd} (see also~\cite{Barreira:2017kxd,Barreira:2019icq,Philcox:2020zyp,Wadekar:2020hax}). In particular, it has been found that the cross-covariance between the power spectrum and the bispectrum is negligible for our scale cuts. For the bispectrum multipoles, we also compute their covariances in the Gaussian tree-level approximation, as detailed in Appendix~\ref{sec:gauss}. Note that the correlation between various multipoles appears already in this approximation (similar to the correlation between different $P_\ell$ multipoles), though we ignore the correlation between the bispectrum multipoles and the power spectrum, as before. Based on the results of~\cite{Ivanov:2021kcd}, this approximation is adequate for our choice of $\kmax^B$.

\subsection{Nseries}
The second type of simulation data we consider is the Nseries mock suite~\cite{BOSS:2016wmc,eBOSS:2020yzd} (see also~\cite{Hahn:2016kiy,Hand:2017ilm}). This suite consists
of 84 pseudo-independent realizations of the BOSS-like halo occupation distribution-based galaxies, covering a cumulative effective volume of, approximately,\footnote{This value is based on the CMASS NGC effective sky area and redshift range given in~\cite{Reid:2015gra}.} $235$~($h^{-1}$Gpc)$^3$. 
The Nseries mocks include all necessary observational effects present in the actual BOSS CMASS sample: the redshift distribution, fiber collisions, and the survey window function. 
As such, these mocks are appropriate to test our window-free estimator, as well as our galaxy clustering model. These mocks were used for validating the official BOSS DR12 data analysis pipeline. 

The effective redshift of the Nseries mocks is $z_{\rm eff}=0.55$ and we analyze the same dataset as in~\eqref{eq:datav} but with $\kmax^{P_\ell}=0.2~\hMpc$,
and $\kmin^{Q_0}=0.2~\hMpc$, consistent with the analysis of Ref.~\cite{Philcox:2021kcw}. The power spectrum and bispectrum multipoles are measured with the unwindowed estimator described in \S\ref{sec:estim}. This uses $100$ Monte Carlo realizations to compute the Fisher matrix and one-point terms. For the pixel weighting, we assume the FKP limit $\mathsf{H}^{-1}\to \delta_{\rm D}(\vr_i-\vr_j)n^{-1}(\vr)[1+n(\vr)P_{\rm FKP}]^{-1}$ for $P_{\rm FKP}=10^4h^{3}\mathrm{Mpc}^{-3}$, with the window function $n(\vr)$ computed from the survey mask and redshift distribution. Our initial bispectra are computed with $\kmax^B=0.11~\hMpc$ then trimmed to $\kmax^B=0.08~\hMpc$ to minimize window-function-induced correlations with. modes not included in the analysis. In the final data vector, we use $62$ bispectrum bins with $\Delta k = 0.01~\hMpc$ for each multipole.

Here, we assume the likelihood for the dataset to be Gaussian (valid since we limit to quasi-linear scales). Since the window function induces non-negligible correlations between the power spectrum and bispectrum (which enters the covariance but not the mean datavector), we cannot use the analytic approximations described above; instead, we use the empirical covariance extracted from the NGC MultiDark Patchy CMASS mocks~\cite{Kitaura:2015uqa,Rodriguez-Torres:2015vqa}. This set of approximate mocks has a selection function and geometry closely matching that of the BOSS CMASS sample.
We use 2048 mocks in our covariance estimator, which guarantees that the sampling noise is heavily suppressed (though see \citep{Philcox:2020zyp} for compression-based appraoches). We stress that all our consistency checks are carried out on realistic mocks such as PT Challenge and Nseries, which are based on exact N-body simulations. The MultiDark Patchy mocks, which are generated with approximate gravity solvers, are used only to build covariance matrices.

\subsection{BOSS}
Finally, we analyze real clustering data, from the twelfth data release (DR12, 2016) of BOSS~\cite{BOSS:2016wmc}. The data is split into four different chunks depending on the redshift coverage and sky position, denoted NGCz1, SGCz1, NGCz3, and SGCz3, where SGC and NGC refer to South and North Galactic Cap survey regions, and z1$=0.38$ and z3$=0.61$ are the sample effective redshifts. The power spectrum and bispectrum multipoles are computed using the window-free estimator described in \S\ref{sec:estim} (see also \cite{Philcox:2020vbm,Philcox:2021ukg}). We supplement the data vector~\ref{eq:datav} with BAO measurements from the reconstructed power spectrum measurements, condensed into Alcock-Paczynski parameters $\alpha_\parallel$, $\alpha_\perp$. These are extracted for each data chunk as described in Ref.~\cite{Philcox:2020vvt}. The likelihood for the full data vector for each of the four BOSS data samples,  
\be
\label{eq:datav2}
\{P_0,P_2,P_4,Q_0,B_0,B_2,B_4,\alpha_\parallel,\alpha_\perp\}\,, 
\ee
is assumed to be Gaussian, with the empirical covariance obtained from the suite of MultiDark Patchy mocks generated 
separately for each data sample. Note that the bispectrum covariance is very close to the one computed in the Gaussian 
tree-level approximation, \textit{i.e.}\ the window function effects are small when using our window-free estimator (though not guaranteed to be zero).

\subsection{Codes \& Priors}
We evaluate our theoretical predictions for the power spectrum 
and bispectrum with the open source \texttt{CLASS-PT} code~\cite{Chudaykin:2020aoj} (see also~\cite{Chen:2020zjt,DAmico:2020kxu}). MCMC chains are computed with the \texttt{Montepython} code~\cite{Audren:2012wb,Brinckmann:2018cvx}.

Finally, let us discuss priors on nuisance parameters. For the power spectrum is concerned, we adopt the same priors as in previous BOSS EFT full-shape analyses, detailed in Refs.~\cite{Chudaykin:2020aoj,Philcox:2021kcw,Philcox:2022frc} (with conventions described in Appendix D of~\cite{Ivanov:2021kcd}). For the bispectrum nuisance parameters, we assume 
\be
A_{\rm shot}\sim \mathcal{N}(0,1^2)\,,\quad  
B_{\rm shot}\sim \mathcal{N}(1,1^2)\,,\quad 
c_1\sim \mathcal{N}(0,5^2)\,,
\ee
which are motivated by naturalness, which implies that the 
EFT parameters should be $\mathcal{O}(1)$ (after removing their physical scalings).

\section{Tests on Mock Catalogs}
\label{sec:mocks}
In this section we test our analysis pipeline on the realistic mock catalogs described above, starting with the PT Challenge mocks. These cover a huge effective volume, and do not contain survey systematics effects, thereby allowing clear tests of our theory model for the anisotropic bispectrum. After this, we will proceed to the Nseries mock suite, which cover a somewhat smaller volume, are not exactly independent (the 84 mocks in the suite are based on only 7 independent N-body realizations),
but include all necessary observational effects present in the actual data, and are thus analyzed using window-free estimators.

In both cases, we will fit for the cosmological parameters of the minimal $\Lambda$CDM model. These are the Hubble constant $H_0$, the physical dark matter density $\omega_{cdm}$, the primordial power spectrum amplitude $A_s$ and tilt $n_s$. We also consider the derived parameters $\Omega_m$ and $\sigma_8$. The CMB temperature $T_0$ is kept fixed to the FIRAS value~\cite{Aghanim:2018eyx}.\footnote{This parameter is not relevant for the LSS data. We require it here only to convert the measured baryon-to-photon and dark-matter-to-photon ratios into $\omega_b$ and $\omega_{cdm}$~\cite{Ivanov:2020mfr}.} The physical baryon fraction, $\omega_b$, is kept fixed to the true value of the mocks in order to simulate the effect of the $\omega_b$ prior from either Big Bang Nucleosynthesis (BBN)~\cite{Aver:2015iza,Cooke:2017cwo} or the CMB. Finally, the neutrino masses are set to zero, as in the simulations. We will find that our pipeline successfully recovers the input cosmological parameters from both types of mocks in this setup.

\subsection{PT Challenge}
\label{sec:ptc}

\begin{table}[!t]
     \centering
     \scriptsize
     \rowcolors{2}{white}{vlightgray}
   \begin{tabular}{c|cccc} \hline\hline
      \textbf{Dataset} & $\Delta\omega_{\rm cdm}/\omega_{\rm cdm}$ & $\Delta H_0/H_0$ & $\Delta A_s/A_s$ & $\Delta n_s/n_s$ \\\hline
      $P_\ell+Q_0+B_0$ \,&\, $-0.004\pm 0.010$ \,&\, $-0.0007\pm 0.0017$ \,&\, $0.007\pm 0.019$ \,&\, $0.0085\pm 0.0077$\\
      $P_\ell+Q_0+B_\ell$ \,&\, $0.0011\pm 0.0099$ \,&\, $-0.0001\pm 0.0017$ \,&\, $-0.017\pm 0.017$ \,&\,$0.0064\pm 0.0077$\\\hline\hline
      \end{tabular}
      \\\vskip 8pt
      \begin{tabular}{c|cc|ccc} \hline\hline
      \textbf{Dataset} & $\Delta\Omega_m/\Omega_m$ & $\Delta \sigma_8/\sigma_8$ 
      & $\Delta b_1/b_1$ & $\Delta b_2$ & $\Delta b_{\mathcal{G}_2}$\\\hline
      $P_\ell+Q_0+B_0$ \,&\,$-0.0021\pm 0.0068$ \,&\, $0.0040\pm 0.0069$ \,&\,$-0.0026\pm 0.0072$ \,&\, $-0.111\pm 0.079$\,&\, $0.025\pm 0.024$\\
      $P_\ell+Q_0+B_\ell$ \,&\, $0.0011\pm 0.0067$ \,&\, $-0.0056\pm 0.0063$ \,&\, $0.0102\pm 0.0063$ \,&\, $0.053\pm 0.058$\,&\, $0.043\pm 0.022$\\\hline\hline
      \end{tabular}
  \caption{One-dimensional marginalized constraints on 
 cosmology and low-order bias parameters extracted from the PT Challenge dataset. The top table shows directly sampled cosmological parameters whilst the bottom shows derived parameters and biases. In each case, we give results including both the bispectrum monopole and multipoles.}
\label{tab:ptc}
\end{table}

\begin{figure}
    \centering
    \includegraphics[width=\textwidth]{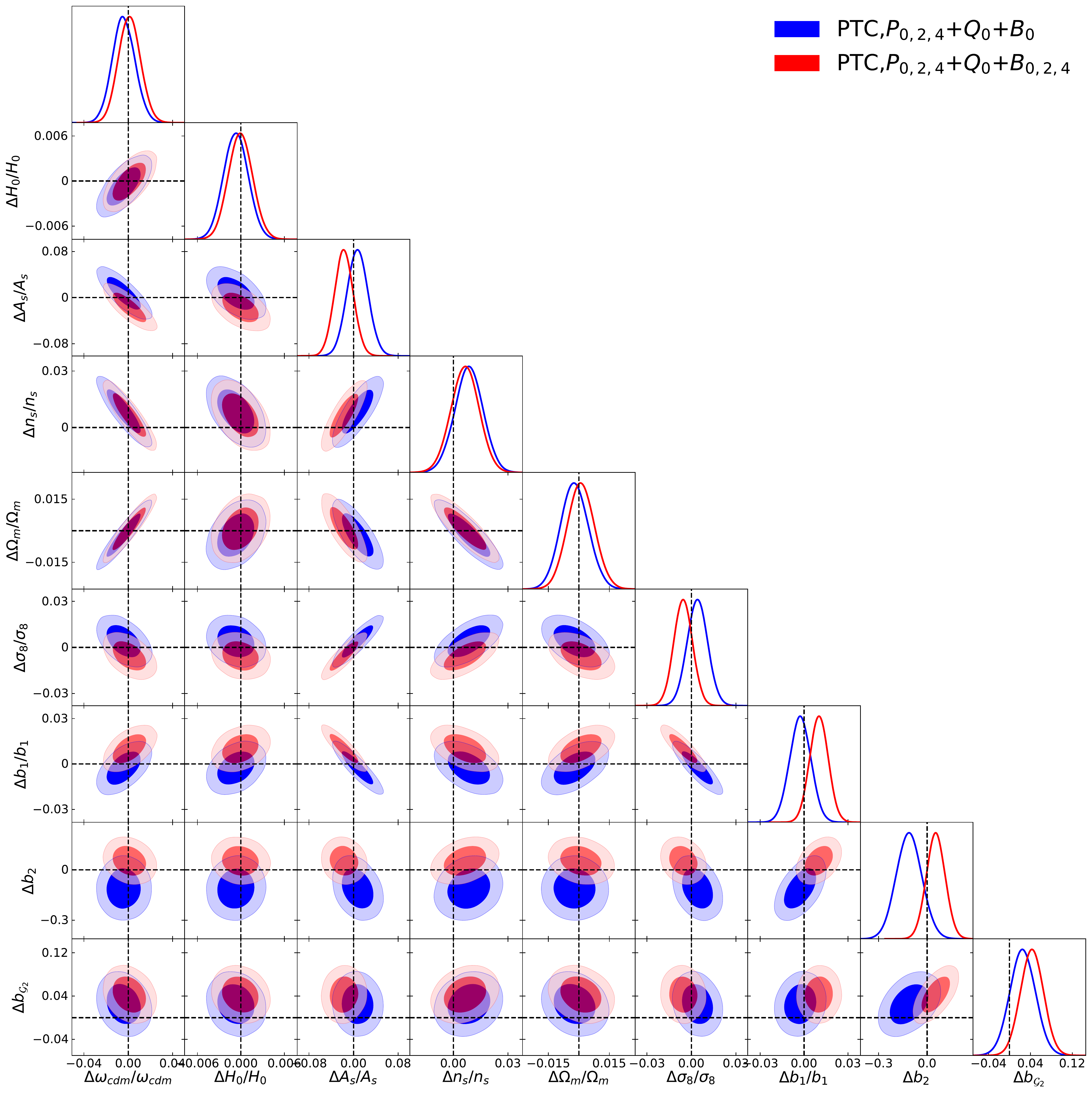}
    \caption{Posteriors on cosmological and main bias parameters extracted from the power spectrum and bispectrum of the PT Challenge simulation. All parameters are normalized to their true values (or their proxy for bias coefficients). The the power spectrum data is the same in both analyses. Blue contours correspond to the bispectrum monopole, whilst those in red result from the addition of the bispectrum quadrupole and hexadecapole moments. We find only small shifts in cosmological parameters, consistent with the errors, and a slight posterior shrinkage.
    }
    \label{fig:ptc_post}
\end{figure}

We begin by considering the likelihood of the PT Challenge power spectrum and bispectrum multipoles. For comparison, we also present results obtained from the bispectrum monopole likelihood, \textit{i.e.}\ that excluding higher-order angular moments. The latter results are equivalent to those present in Ref.~\cite{Ivanov:2021kcd}. The posteriors of cosmological, 
linear and quadratic bias parameters extracted from the PT Challenge simulation data are displayed in Fig.~\ref{fig:ptc_post}, with the one-dimensional marginalized limits given in Tab.~\ref{tab:ptc}. Since the PT challenge is still on-going, the presented cosmological parameters are normalized to their true values that we keep unknown to the reader. A similar logic holds for the linear bias parameter, $b_1$, whose ground truth value is taken from fits to 
the real-space one-loop galaxy power spectrum and bispectrum datasets~\cite{Philcox:2022frc}. For the quadratic bias parameters, we instead display $\Delta b_2 = b_2-b_2^{\rm truth}$, $\Delta b_{\mathcal{G}_2} = b_{\mathcal{G}_2} -b_{\mathcal{G}_2} ^{\rm truth}$, where the ground truth values are adapted from~\cite{Ivanov:2021kcd}.

Looking at Fig.~\ref{fig:ptc_post} and Tab.~\ref{tab:ptc}, we see that our fitting pipeline successfully recovers the cosmological and main nuisance parameters from the PT Challenge data. The second relevant observation is that the addition of the bispectrum multipoles does not have a strong impact on the cosmological parameter recovery. One can notice some $\lesssim 0.5\sigma$ shifts in the posterior means for some cosmological parameters, and a modest shrinking of the errorbars. The largest effect is on $\sigma_8$ (and $b_1$), whose posteriors narrow by $\lesssim 10\%$. In contrast to cosmological parameters, the effect on the quadratic bias parameters is more pronounced, with $b_2$ and $b_{\mathcal{G}_2}$ posteriors shrinking by 30\% and 10\%, respectively.

The best-fitting theory models for the bispectrum multipoles are shown in Fig.~\ref{fig:ptc-plot}. Here, we display the full bispectrum dataset as a function of the triangle index, as well as squeezed and equilateral configurations as functions of relevant wavenumbers of the bin centers. As expected, we find excellent agreement between theory and data for all multipoles considered.
\subsection{Nseries}
\label{sec:nser}

\begin{table}[!t]
     \centering
     \scriptsize
     \rowcolors{2}{white}{vlightgray}
   \begin{tabular}{c|cccc} \hline\hline
      \textbf{Dataset} & $\omega_{\rm cdm}$ & $H_0$ & $\ln\left(10^{10}A_s\right)$ & $n_s$ \\\hline
      $P_\ell+Q_0+B_0$ \,&\, $0.1158\pm 0.0021$ \,&\, $70.09\pm 0.21$ \,&\,$3.103\pm 0.033$ \,&\, $0.986\pm 0.014$\\
      $P_\ell+Q_0+B_\ell$ \,&\, $0.1153\pm 0.0020$ \,&\, $70.09\pm 0.20$ \,&\,$3.114\pm 0.032$\,&\,$0.986\pm 0.013$\\
      $P_\ell+Q_0+B_\ell, V_{\rm BOSS}$\,&\,$0.1198^{+0.0092}_{-0.012} $ \,&\, $70.4^{+1.0}_{-1.2}$ \,&\, $2.99\pm 0.16$ \,&\, $0.959\pm 0.067$\\\hline\hline
      \end{tabular}
      \\\vskip 8pt
      \begin{tabular}{c|cc|ccc} \hline\hline
      \textbf{Dataset} & $\Omega_m$ & $\sigma_8$ 
      & $b_1$ & $\Delta b_2$ & $\Delta b_{\mathcal{G}_2}$\\\hline
      $P_\ell+Q_0+B_0$ \,&\,$0.2825\pm 0.0032$ \,&\, $0.838\pm 0.010$\,&\, $1.980\pm 0.024$ \,&\, $-0.27\pm 0.11$\,&\,$-0.252\pm 0.050$\\
      $P_\ell+Q_0+B_\ell$ \,&\, $0.2815\pm 0.0031$ \,&\, $0.8407\pm 0.0097$ \,&\, $1.968\pm 0.023$ \,&\, $-0.312\pm 0.091$ \,&\, $-0.207\pm 0.045$\\
      $P_\ell+Q_0+B_\ell, V_{\rm BOSS}$\,&\,$0.288^{+0.015}_{-0.018}$\,&\,$0.801^{+0.043}_{-0.052}$\,&\,$2.07\pm 0.12$\,&\, $-0.07^{+0.41}_{-0.47}$\,&\,$-0.16\pm 0.22$\\\hline\hline
      \end{tabular}
  \caption{Marginalized constraints on cosmology and low-order bias parameters extracted from the Nseries dataset. As in Tab.\,\ref{tab:ptc}, we show sampled cosmological parameters in the first table and derived parameters and low-order biases in the second. The first and second row shows results for the 84 Nseries mocks with the single mock covariance divided by 84 to match the true cumulative volume, whilst the third row gives results for the same mean data vector, but with the covariance rescaled to match the BOSS volume $V_{\rm BOSS}\approx 6~h^{-3}$Gpc$^3$, thus probing prior-volume effects. The true cosmological parameter values are given by $\omega_{cdm}=0.11711$, $H_0=70~\mathrm{km}\,\mathrm{s}^{-1}\mathrm{Mpc}^{-1}$, $n_s=0.96$, $\ln (10^{10}A_s)=3.0657$, $\Omega_m=0.286$, and $\sigma_8 = 0.82$.}
\label{tab:nseries}
\end{table}

\begin{figure}
    \centering
    \includegraphics[width=\textwidth]{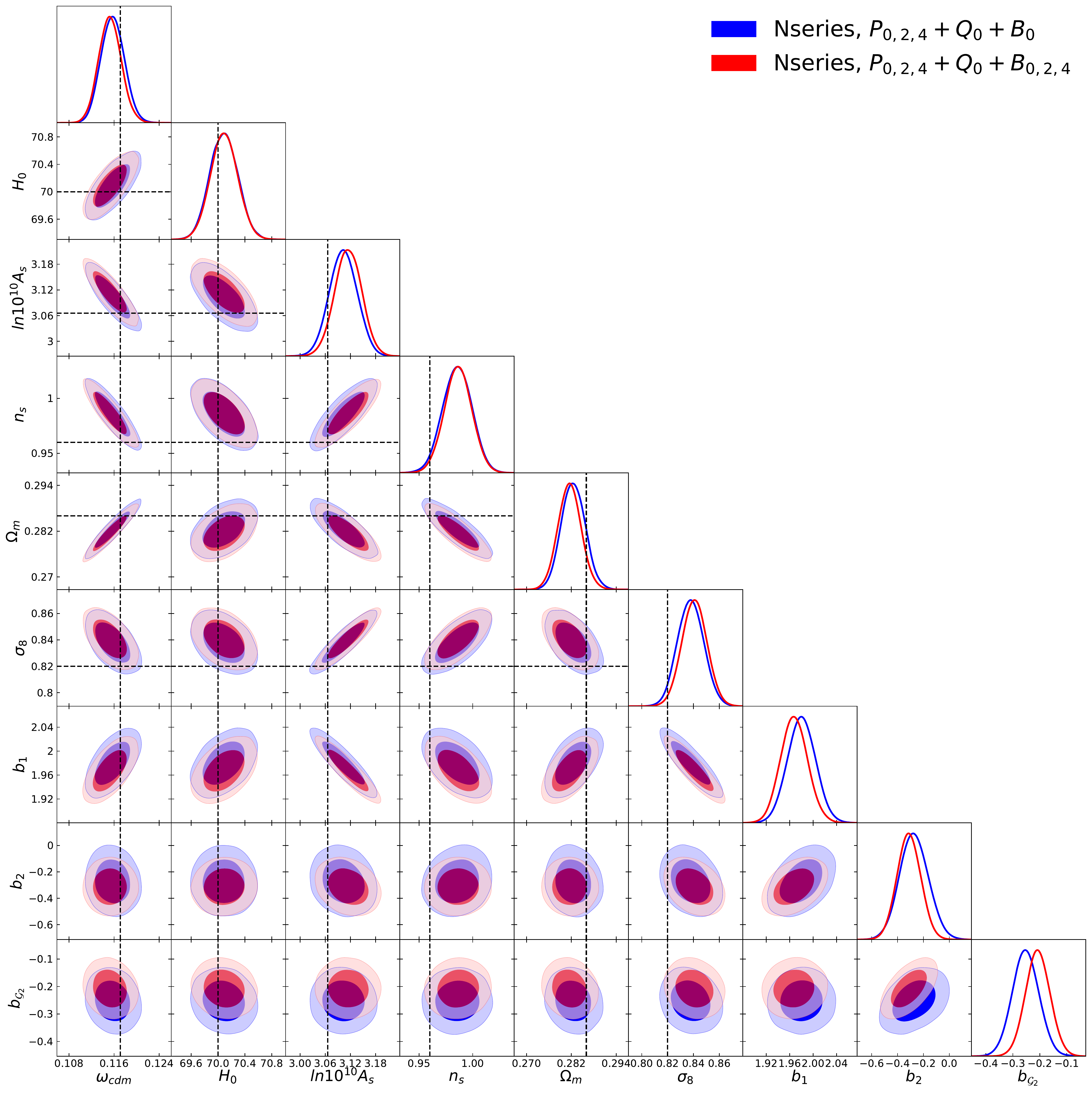}
    \caption{As Fig.~\ref{fig:ptc_post}, but for the Nseries dataset. We give one-dimensional posteriors in Tab.\,\ref{tab:nseries}.}
    \label{fig:nser_post}
\end{figure}

\begin{figure}
    \centering
    \includegraphics[width=\textwidth]{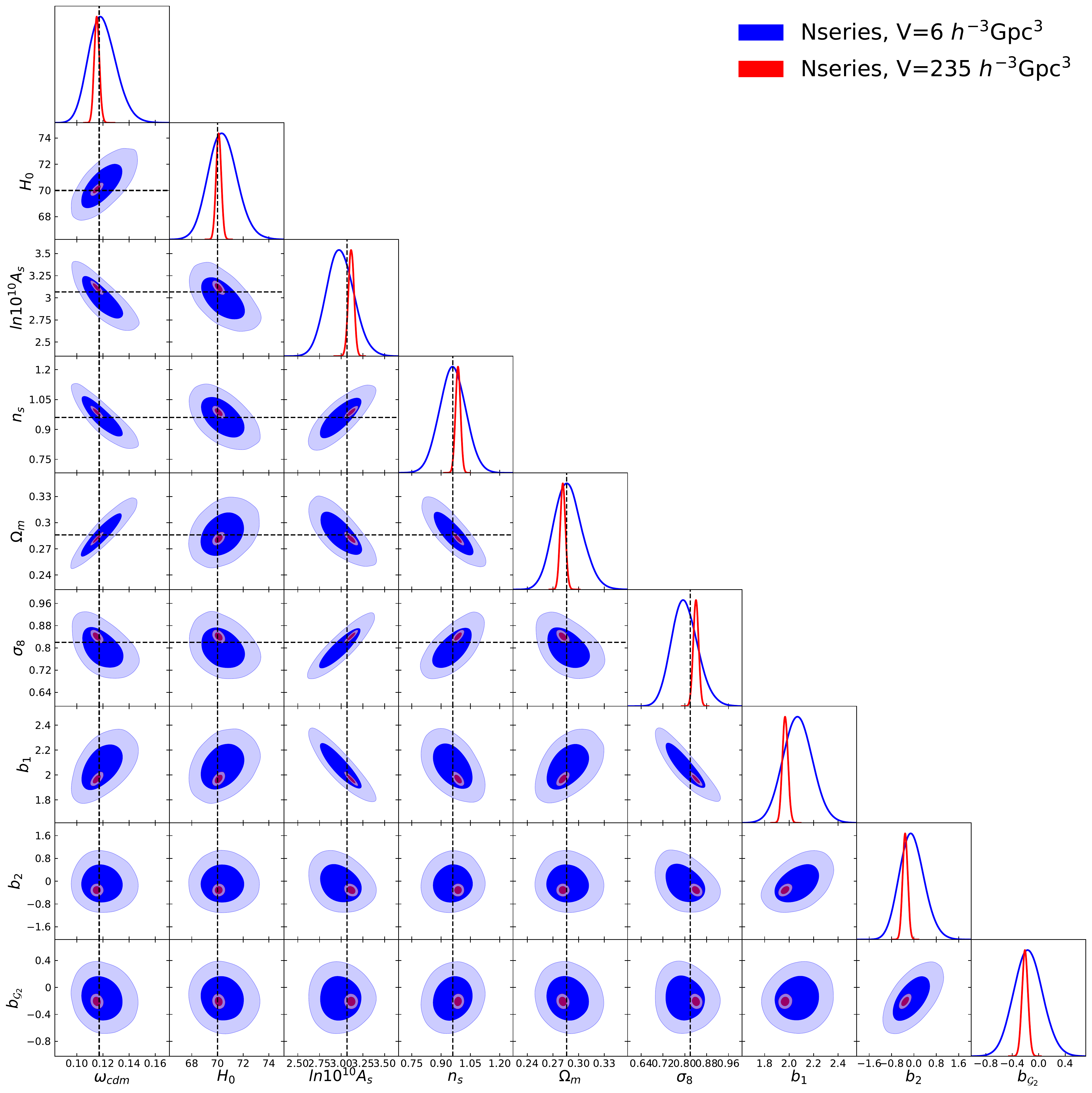}
    \caption{As Fig.~\ref{fig:nser_post}, but comparing constraints on Nseries power constraints between analyses using a covariance matching the entire Nseries volume ($\approx 235~h^{-3}$Gpc$^3$) and that of BOSS ($\approx 6~h^{-3}$Gpc$^3$). Whilst there is some evidence prior volume effects (such as in $\sigma_8$), the corresponding shifts are subdominant compared to the errorbars.}
    \label{fig:nser_postv6}
\end{figure}

Let us now move to the Nseries mocks. Our results for this dataset are shown in Fig.~\ref{fig:nser_post} and in Tab.~\ref{tab:nseries}. As before, we observe that our pipeline successfully recovers the input cosmological parameters used in the simulation, thus validating the window-free estimators of \S\ref{sec:estim}. Once again, the bispectrum multipoles have the strongest impact on the $\sigma_8$ posteriors, which are $\approx 5\%$ narrower than those from the bispectrum monopole likelihood. In addition, the $b_2$ and $b_{\mathcal{G}_2}$ posteriors shrink by 20\% and 5\% respectively.

Overall, the improvements obtained found for the Nseries mocks are somewhat smaller than the improvements seen in the PT Challenge case. We believe that this difference is caused by the Gaussian tree-level approximation for the bispectrum likelihood used in the PT Challenge case. For the Nseries dataset we use the full covariance extracted from mocks, which is more reliable than the naive Gaussian approximation, and accounts for mode-coupling induced by non-linear clustering. 

All in all, we conclude that our pipeline is capable of unbiased recovery of cosmological parameters from the actual data. We have demonstrated that the theory model works well on both high-fidelity periodic box data, as well as on mocks with realistic survey geometry and observational effects. Our tests on Nseries mocks additionally imply that our window-free estimator robustly recovers the true bispectrum of anisotropic galaxy clustering.
 
It is also important to estimate the importance of effects arising from our choice of Gaussian priors, since these may shift the posteriors of a Bayesian analysis away from the true values~\cite{Ivanov:2019pdj,Chudaykin:2020ghx,Philcox:2021kcw}.
To this end we repeat our Nseries analysis, but using a covariance corresponding to the BOSS cumulative volume 6~($h^{-1}$Gpc)$^3$, with the datavector still given by a mean over 84 Nseries realizations. This set-up simulates the situation where we analyze separately 84 (semi)-independent realizations (with the BOSS covariance each), and average over our results instead of combining them (changing the ratio of likelihood to prior relative to the above test). In what follows 
we will call the covariance corresponding to the true cumulative simulation volume ``true covariance,''
and the covaraince rescaled to match the BOSS volume as the ``BOSS covariance.''

The outcome of this analysis is shown in Fig.~\ref{fig:nser_postv6} and Tab.~\ref{tab:nseries}. 
We see that the mean value of $\sigma_8$ from the analysis with the BOSS covariance is lower than that from the analysis with the true covariance of 84 realizations (emulating a much larger survey). Since both likelihoods are identical except for an overall multiplication of the covariance, we interpret the observed shifts as a result of prior volume (marginalization) effects. The maximum-likelihood (but not maximum a posteriori) value of $\sigma_8$ remains the same in both analyses as it is not affected by the rescaling of the covariance matrix. Let us denote the one-dimensional marginalized errorbar on $\sigma_8$ from the BOSS analysis as $\sigma_{\rm BOSS}$. From the true-covariance results, we find that the best-fit is biased up by $\approx 2\%$ with respect to the true value of $\sigma_8$, or by $0.4\sigma_{\rm BOSS}$. This may be interpreted as a true systematic error, although it is small enough that we cannot robustly rule out the possibility that it is a statistical fluctuation. The average mean value resulting from the BOSS covariance analysis is shifted by $0.4\sigma_{\rm BOSS}$ away from the actual input value and $0.8\sigma_{\rm BOSS}$ from the best-fit (which nearly coincides with the mean of the analysis with the true covariance). However, the actual metric we are interested in is the shift of the average mean with respect to the true fiducial value, which is well below the errorbars. We thus conclude that the prior volume effects are not significant for our analysis.

\section{Analysis of the BOSS data}\label{sec:boss}

We now present parameter constraints from the BOSS DR12 dataset and estimate the information content of the galaxy bispectrum multipoles, see table~\ref{tab:snrt}. The full constraint table including the nuisance parameters is presented in Appendix~\ref{sec:fullconst}. We begin by considering the actual measurements from the data, obtained using the unwindowed estimators of \S\ref{sec:estim}. In Fig.~\ref{fig:boss_bisp_data} we present the window-free galaxy bispectrum multipoles extracted from the NGCz3 data chunk. {Our first relevant observation is that only the monopole moment carries a high signal, \textit{i.e.} it is detected at $\approx 20\sigma$ . The quadrupole is detected at a relatively lower significance, $\approx 5\sigma$, whilst the hexadecapole contribution is not detected at all. }

{Although the detection significance of the large-scale bispectrum multipoles is lower than that of the monopole,} it does not mean that they are devoid of cosmological information. Indeed, what is relevant for actual cosmological constraints is not the signal-to-noise \textit{per se}, but the amplitude of Fisher derivatives. In other words, the bispectrum multipoles may still be useful, e.g. in the breaking of certain parameter degeneracies. To check this, we proceed now to the actual MCMC analysis of our likelihood containing the bispectrum multipole moments. In this vein, we will compare the parameter constraints from our likelihood including the bispectrum multipoles to that containing only the bispectrum monopole. 

We begin with the \textit{Planck}-independent $\Lambda$CDM analysis, \textit{i.e.}\ that with free tilt $n_s$. Our results are displayed in Fig~\ref{fig:boss_post_ns} and Tab.~\ref{tab:boss}, showing results for the cosmological parameters only. We find that the bispectrum multipoles narrow the posteriors only marginally, by $\lesssim 10\%$, with the largest effect on $n_s$, whose errorbar has shrunk by 10\%. We also find a (broadly insignificant) $\approx 0.2\sigma$ upward shift in the $\Omega_m-\sigma_8$ plane. 

Imposing the \textit{Planck} prior on $n_s$ does not qualitatively change the situation: we observe marginal improvements on all cosmological parameters in addition to a small upward shift of the $\Omega_m-\sigma_8$ posterior, see Fig.~\ref{fig:boss_post_ns}. To investigate the origin of this shift, we have repeated our analysis with the same data, but with a covariance matrix in which we have artificially removed the correlation between $P_\ell$ and $B_\ell$ data sets. In this case, we find that the mean values do not noticeably shift with respect to the $P_\ell+Q_0+$BAO+$B_0$ analysis. In particular, we find $\Omega_m=0.3156_{-0.0099}^{+0.0094}$, $H_0=68.21_{-0.86}^{+0.85}~\mathrm{km}\,\mathrm{s}^{-1}\mathrm{Mpc}^{-1}$, $\sigma_8 = 0.7262_{-0.036}^{+0.032}$ (cf.\,Tab.\,~\ref{tab:boss}). Further investigation reveals that certain elements of the $P_\ell-B_\ell$ correlation matrix are enhanced relative to the linear theory Gaussian approximation, which may be a result of the non-trivial survey window function geometry, or a limitation of the (approximate) Patchy simulations. Our study suggests that it is this correlation that produces 
the apparent $\sim 0.5\sigma$ shift in the $\Omega_m-\sigma_8$ plane. We leave further investigation of this effect for future work.

We note that the addition of the bispectrum multipoles leads to a significantly more Gaussian posterior for $\sigma_8$: we find $\sigma_8 = 0.736\pm 0.033$. In addition, our result is now in greater harmony with the \textit{Planck} 2018 $\Lambda$CDM constraint $\sigma_8=0.811\pm 0.006$~\cite{Aghanim:2018eyx}. We close by noting that our final $\sigma_8$ result is nominally the strongest of all previously reported full-shape measurements based on the EFTofLSS.

\begin{figure}
    \centering
    \includegraphics[width=\textwidth]{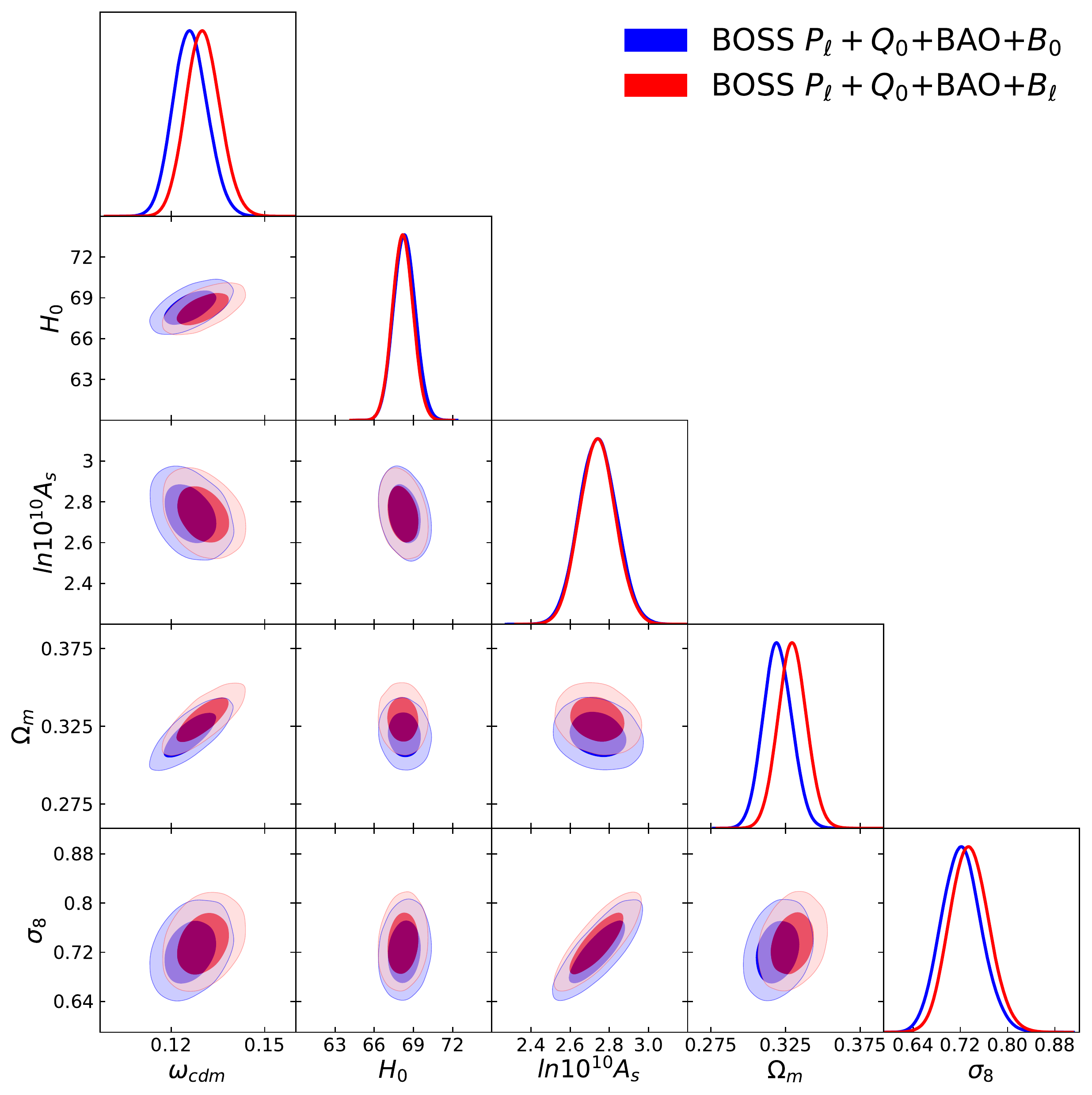}
    \caption{As Fig.~\ref{fig:boss_post_ns}, but for an analysis with $n_s$ fixed to 
    the \textit{Planck} best-fit value. 
    }
    \label{fig:boss_post}
\end{figure}

\section{Discussion and Conclusions}
\label{sec:disc}

In this work we have performed a cosmological analysis of the BOSS galaxy power spectrum and bispectrum, that for the first time self-consistently includes the large-scale ($k<0.08~\hMpc$) bispectrum quadrupole and hexadecapole. The BOSS bispectrum moments are extracted using a novel window-free estimator, derived within a maximum-likelihood formalism. This allows us to reconstruct the underlying anisotropic bispectrum (\textit{i.e.}\ that unconvolved with the survey window function), and significantly simplifies consequent data analyses, since our measurements can be directly 
compared with theory.

Our pipeline has been validated using two sets of mocks,
which have established that the method's systematic errors are significantly below the statistical ones. In particular, we have analyzed the multipole moments of the PT Challenge simulation suite, which covers a gigantic volume of $566$ $h^{-3}$Gpc$^3$. We obtained an excellent fit of theory and simulation, and were able to recover unbiased true cosmological parameters in all our tests. This implies that our pipeline matches the precision requirements of future surveys such as DESI~\cite{Aghamousa:2016zmz} and Euclid~\cite{Laureijs:2011gra,Amendola:2016saw,Sprenger:2018tdb}. 

Assuming the minimal $\Lambda$CDM model, we have found that the inclusion of the higher galaxy bispectrum multipoles narrow the constraints only moderately (with typical improvements for the one-dimensional posterior distributions at the level of $(5-10)\%$). The main reason for this is that the higher bispectrum multipoles contain much less signal and much larger noise than the large-scale power spectrum and bispectrum monopole. 
This is consistent with previous work~\cite{DAmico:2022osl}, which showed that the addition of the large-scale BOSS bispectrum quadrupole data only improved the constraint on $\Omega_m$
by $\sim 10\%$. Nevertheless, taking into account the information in the bispectrum monopole as well, these results imply that the total improvement from the 
redshift-space bispectrum compared the power spectrum alone can be significant, and as large as $\sim 20\%$.
It is also worth commenting on Ref.~\cite{Rizzo:2022lmh}, which found some noticeable improvement on $f\sigma_8(z)$ from the bispectrum multipoles. Our analysis is principally different from~\cite{Rizzo:2022lmh} in that we analyze the bispectrum multipoles in conjunction with the power spectrum and BAO data. Our results suggest that for this type of analysis the $f\sigma_8(z)$ constrains are largely dominated by the power spectrum likelihood, and the impact of the bispectrum multipoles is somewhat modest. The information gain may be bigger if one pushes the analysis to smaller scales, which would require either a one-loop perturbative model~\cite{Philcox:2022frc,DAmico:2022osl} or a simulation-based emulator~\cite{Hahn:2019zob,Hahn:2020lou}.
We plan to explore the first option in the future. 

Another important caveat is that our analysis has been performed only for the minimal $\Lambda$CDM model. One might hope that the relative improvement from the bispectrum multipoles is larger for extended cosmological models (as observed for the power spectrum multipoles, e.g., \cite{Ivanov:2020ril,DAmico:2020tty,DAmico:2020ods,Chudaykin:2020ghx}, see also~\cite{Tsedrik:2022cri} for the bispectrum quadrupole in the context of interacting dark energy models). In particular, the bispectrum is a sensitive probe of early universe physics~\cite{Sefusatti:2007ih,Sefusatti:2009qh,MoradinezhadDizgah:2018ssw,MoradinezhadDizgah:2020whw,Cabass:2022wjy,Cabass:2022ymb,Cabass:2022epm,Green:2022bre} and hypothetical violations of the equivalence principle~\cite{Creminelli:2013nua} that are motivated, for example, by Lorentz-violating dark matter models~\cite{Blas:2012vn,Audren:2014hza}, 
long-range forces in the dark sector~\cite{Archidiacono:2022iuu} or non-trivial dark energy theories~\cite{Crisostomi:2019vhj,Lewandowski:2019txi}. In addition, it would be interesting to understand if the bispectrum multipoles can sharpen full-shape constraints on other non-minimal dark matter models~\cite{Lague:2021frh,Xu:2021rwg,Nunes:2022bhn,Rubira:2022xhb,Rogers:2023ezo,He:2023dbn}, additional long-range interactions in the dark sector~\cite{Archidiacono:2022iuu} or some non-minimal dark energy theories~\cite{Crisostomi:2019vhj,Lewandowski:2019txi}. We leave the exploration of these interesting possibilities to future work. 

\paragraph{Acknowledgments} 
{\footnotesize
We would like to thank Kazuyuki Akitsu and Shi-Fan Chen for useful discussions. 
The work of MMI has been supported by NASA through the NASA Hubble Fellowship grant \#HST-HF2-51483.001-A awarded by the Space Telescope Science Institute, which is operated by the Association of Universities for Research in Astronomy, Incorporated, under NASA contract NAS5-26555. OHEP is a Junior Fellow of the Simons Society of Fellows and thanks the Simons Foundation for support. OHEP also acknowledges the Institute for Advanced Study for their hospitality and venison selection. GC acknowledges support from the Institute for Advanced Study. MZ is supported by the Canadian Institute for Advanced Research (CIFAR) program on Gravity and the Extreme
Universe and the Simons Foundation Modern Inflationary Cosmology initiative. This work was supported in part by MEXT/JSPS KAKENHI Grant Number JP19H00677, JP20H05861, JP21H01081 and JP22K03634. We also acknowledge financial support from Japan Science and Technology Agency (JST) AIP Acceleration Research Grant Number JP20317829.

The simulation data analysis was performed partly on Cray XC50 at Center for Computational Astrophysics, National Astronomical Observatory of Japan. Data analysis was partly performed on the Helios cluster at the Institute for Advanced Study, Princeton, and partly using the Princeton Research Computing resources at Princeton University, which is a consortium of groups led by the Princeton Institute for Computational Science and Engineering (PICSciE) and the Office of Information Technology's Research Computing Division.
}
\appendix 

\section{Gaussian Covariance for Bispectrum Multipoles}
\label{sec:gauss}

In this section we present analytic formulae for the 
Gaussian tree-level bispectrum multipole covariance
in the narrow bin approximation, $\Delta k\ll k$~\cite{Scoccimarro:2015bla}. As in \eqref{eq: Bmult-scocc}, the ideal estimator for the bispectrum multipole $\ell$ is given by
\be
\label{eq:realest}
\begin{split}
& \hat B_{\ell}(k_1,k_2,k_3) = \frac{(2\ell+1)}{N_T^{123}}\prod_{i=1}^3\int_{\vk_1\vk_2\vk_3} (2\pi)^3\delta^{(3)}_D(\k_{123})
\delta_g(\k_1)\delta_g(\k_2)\delta_g(\k_3)\mathcal{L}_{\ell}(\hat{\z}\cdot \hat{\k}_3)\,,
\end{split}
\ee 
where $N^{123}_T=8\pi^2 k_1k_2k_3\Delta k^3 V^2/(2\pi)^6$ (in the thin-bin limit), $V=(2\pi)^3k_f^{-3}$, and $k_f$ is the fundamental wavenumber. At linear order, the galaxy density can be written $\delta_g(\k)=\delta(\k)(1+\beta \mu^2)+\epsilon$~\cite{Kaiser:1987qv}, where $\beta\equiv f/b_1$ and $\epsilon$
is the stochastic density component, whose power spectrum 
we assume to be equal to $\bar n^{-1}$.

Using Eq.~\eqref{eq:realest}, we obtain the bispectrum covariance between triangle configurations $T$ and $T'$,
\be 
\begin{split}
& \langle \hat B_{\ell}(k_1,k_2,k_3)\hat B_{\ell'}(k'_1,k'_2,k'_3)\rangle  =C^{\ell \ell'}_{TT'} =(2\ell+1)(2\ell'+1)\frac{(2\pi)^3\pi}{k_1k_2k_3\Delta k^3 V}\delta_{TT'}\\
& \times
\Bigg(F_{\ell\ell'}(k_1,k_2,k_3)\prod_{i=1}^3 
P_{11}(k_i)+
\frac{1}{\bar n}
\sum_{i<j, i=1}^{j=3}
P_{11}(k_i)P_{11}(k_j)G_{\ell\ell'}(k_i,k_j)\\
&+ 
\frac{1}{\bar n^2}
\sum_{n=1}^3
P_{11}(k_n)H_{\ell\ell'}(k_n) +
J_{\ell\ell'} \frac{1}{\bar n^3}
\Bigg)\,,\\
\end{split}
\ee
where the multipole-dependent form factors for the purely 
continuous part are given by (defining writing the $\mu_1$,$\mu_2$ angles in terms of $\mu\equiv\mu_3$ and $\phi$)
\be
\begin{split}
 F^{\text{ general}}_{\ell\ell'}=&\int_0^{2\pi} \frac{d\phi}{2\pi}\int_0^1d\mu~(1+\beta \mu^2)^2(1+\beta \mu_1(\mu,\phi)^2)^2(1+\beta \mu_2(\mu,\phi)^2)^2 \mathcal{L}_{\ell}(\mu)\mathcal{L}_{\ell'}(\mu)\,,\\
 F^{\text{isosceles I}}_{\ell\ell'}=&2F^{\text{ general}}_{\ell\ell'}\,,\\
 F^{\text{isosceles II}}_{\ell\ell'}=&\int_0^{2\pi} \frac{d\phi}{2\pi}\int_0^1d\mu~(1+\beta \mu^2)^2(1+\beta \mu_1(\mu,\phi)^2)^2(1+\beta \mu_2(\mu,\phi)^2)^2 \\
\times
&\mathcal{L}_{\ell}(\mu)(\mathcal{L}_{\ell'}(\mu)+\mathcal{L}_{\ell'}(\mu_1))\,,\\
F^{\rm equilateral}_{\ell\ell'}=&\int_0^{2\pi} \frac{d\phi}{2\pi}\int_0^1d\mu~(1+\beta \mu^2)^2(1+\beta \mu_1(\mu,\phi)^2)^2(1+\beta \mu_2(\mu,\phi)^2)^2 \\
\times
&2\mathcal{L}_{\ell}(\mu)(\mathcal{L}_{\ell'}(\mu)+\mathcal{L}_{\ell'}(\mu_1)+\mathcal{L}_{\ell'}(\mu_2))\,,
\end{split}
\ee
the continuous $\times$ stochastic terms are (assuming $i=1,2,~j=2,3,j>i$):
\be
\begin{split}
 &G^{\text{ general}}_{\ell\ell'}=\int_0^{2\pi} 
 \frac{d\phi}{2\pi}\int_0^1d\mu~(1+\beta \mu_i(\mu,\phi)^2)^2(1+\beta \mu_j(\mu,\phi)^2)^2 \mathcal{L}_{\ell}(\mu)\mathcal{L}_{\ell'}(\mu)\,,\\
 &G^{\text{isosceles I}}_{\ell\ell'}=2G^{\text{ general}}_{\ell\ell'}\,,\\
 &G^{\text{isosceles II}}_{\ell\ell'}=\int_0^{2\pi} \frac{d\phi}{2\pi}\int_0^1d\mu~(1+\beta \mu_i(\mu,\phi)^2)^2(1+\beta \mu_j(\mu,\phi)^2)^2 \mathcal{L}_{\ell}(\mu)(\mathcal{L}_{\ell'}(\mu)+\mathcal{L}_{\ell'}(\mu_1))\,,\\
&G^{\rm equilateral}_{\ell\ell'}=2\int_0^{2\pi} \frac{d\phi}{2\pi}\int_0^1d\mu~(1+\beta \mu_i(\mu,\phi)^2)^2(1+\beta \mu_j(\mu,\phi)^2)^2 \mathcal{L}_{\ell}(\mu)(\mathcal{L}_{\ell'}(\mu)+\mathcal{L}_{\ell'}(\mu_1)+\mathcal{L}_{\ell'}(\mu_2))\,,
\end{split}
\ee
and $(n=1,2,3)$
\be
\begin{split}
 &H^{\text{ general}}_{\ell\ell'}=\int_0^{2\pi} 
 \frac{d\phi}{2\pi}\int_0^1d\mu~(1+\beta \mu_n(\mu,\phi)^2)^2 \mathcal{L}_{\ell}(\mu)\mathcal{L}_{\ell'}(\mu)\,,\\
 &H^{\text{isosceles I}}_{\ell\ell'}=2H^{\text{ general}}_{\ell\ell'}\,,\\
 &H^{\text{isosceles II}}_{\ell\ell'}=\int_0^{2\pi} \frac{d\phi}{2\pi}\int_0^1d\mu~(1+\beta \mu_n(\mu,\phi)^2)^2\mathcal{L}_{\ell}(\mu)(\mathcal{L}_{\ell'}(\mu)+\mathcal{L}_{\ell'}(\mu_1)\,,\\
&H^{\rm equilateral}_{\ell\ell'}=2\int_0^{2\pi} \frac{d\phi}{2\pi}\int_0^1d\mu~(1+\beta \mu_n(\mu,\phi)^2)^2 \mathcal{L}_{\ell}(\mu)(\mathcal{L}_{\ell'}(\mu)+\mathcal{L}_{\ell'}(\mu_1)+\mathcal{L}_{\ell'}(\mu_2))\,,
\end{split}
\ee
whilst the purely stochastic contributions are 
\be
\begin{split}
 &J^{\text{ general}}_{\ell\ell'}=\int_0^{2\pi} 
 \frac{d\phi}{2\pi}\int_0^1d\mu~\mathcal{L}_{\ell}(\mu)\mathcal{L}_{\ell'}(\mu)\propto \delta_{\ell \ell'}\,,\\
 &J^{\text{isosceles I}}_{\ell\ell'}=2J^{\text{ general}}_{\ell\ell'}\,,\\
 &J^{\text{isosceles II}}_{\ell\ell'}=\int_0^{2\pi} \frac{d\phi}{2\pi}\int_0^1d\mu~\mathcal{L}_{\ell}(\mu)(\mathcal{L}_{\ell'}(\mu)+\mathcal{L}_{\ell'}(\mu_1))\,,\\
&J^{\rm equilateral}_{\ell\ell'}=2\int_0^{2\pi} \frac{d\phi}{2\pi}\int_0^1d\mu~\mathcal{L}_{\ell}(\mu)(\mathcal{L}_{\ell'}(\mu)+\mathcal{L}_{\ell'}(\mu_1)+\mathcal{L}_{\ell'}(\mu_2))\,,
\end{split}
\ee
where we recall that we have chosen $k_1\leq k_2\leq k_3$ without loss of generality and defined
\be
\begin{split}
\text{general:}&~k_1<k_2<k_3\,,\\
\text{equilateral:}&~k_1=k_2=k_3\,,\\
\text{isosceles I:}&~k_1=k_2<k_3\,,\\
\text{isosceles II:}&~k_1<k_2=k_3\,.
\end{split} 
\ee

In the absence of the AP distortions, the integrals in the form factors $F,G,H,J$ can be evaluated analytically. Since the AP effect is typically quite weak, $\mathcal{O}(1\%)$, we ignore it when evaluating the covariance matrix. Finally, we note that we use the Gaussian covariance for bispectrum multipoles only in the analysis of the PT challenge data. For the Nseries mocks and the BOSS data we use the covariance estimated from the Multi-Dark Patchy mocks, allowing us to incorporate the effects of window functions and non-linear gravity. 

\section{Full constraints and parameter tables}
\label{sec:fullconst}

In Tabs.\,\ref{tab: all-constraints-free-ns}\,\&\,\ref{tab: all-constraints-fix-ns}, we display one-dimensional marginalized constraints on cosmological and nuisance parameters for the $\Lambda$CDM fits to the BOSS data with, respectively, free $n_s$ and $n_s$ fixed to the \textit{Planck} best-fit value. In the left and right panels we show results before and after adding the 
bispectrum multipoles.

\begin{table}
    \centering
   \rowcolors{2}{white}{vlightgray}
   \scriptsize
  \begin{tabular}{|c|cccc|cccc|}
 \hline
 & \multicolumn{4}{c|}{$\bm{P_\ell + Q_0 + \mathrm{BAO} + B_0}$} & \multicolumn{4}{c|}{$\bm{P_\ell + Q_0 + \mathrm{BAO} + B_\ell}$}\\\hline
\quad \textbf{Parameter}\quad\quad & best-fit & mean$\,\pm\,\sigma$ & \quad 95\% lower \quad & \quad 95\% upper \quad & best-fit & mean$\,\pm\,\sigma$ & \quad 95\% lower \quad & \quad 95\% upper \quad \\ \hline
$\omega_{cdm }$ &$0.1348$ & $0.1398_{-0.013}^{+0.01}$ & $0.1168$ & $0.1649$ &$0.1405$ & $0.1444_{-0.013}^{+0.01}$ & $0.1215$ & $0.1684$ \\
$h$ &$0.6903$ & $0.6928_{-0.012}^{+0.011}$ & $0.67$ & $0.7159$ &$0.6923$ & $0.6919_{-0.011}^{+0.01}$ & $0.6704$ & $0.7139$ \\
$\mathrm{ln}\left(10^{10}A_{s}\right)$ &$2.69$ & $2.598_{-0.14}^{+0.13}$ & $2.335$ & $2.868$ &$2.626$ & $2.602_{-0.13}^{+0.12}$ & $2.354$ & $2.858$ \\
$n_{s }$ &$0.8915$ & $0.8724_{-0.065}^{+0.069}$ & $0.7371$ & $1.007$ & $0.8778$ & $0.869_{-0.061}^{+0.062}$ & $0.7471$ & $0.9924$ \\
$b^{(1)}_{1 }$ &$2.317$ & $2.419_{-0.13}^{+0.13}$ & $2.163$ & $2.682$ &$2.39$ & $2.418_{-0.13}^{+0.12}$ & $2.166$ & $2.667$ \\
$b^{(1)}_{2 }$ &$0.02127$ & $0.4025_{-0.78}^{+0.71}$ & $-1.08$ & $1.894$ &$0.1334$ & $0.3783_{-0.76}^{+0.69}$ & $-1.046$ & $1.839$ \\
$b^{(1)}_{{\mathcal{G}_2} }$ &$-0.393$ & $-0.3541_{-0.36}^{+0.37}$ & $-1.09$ & $0.3627$ &$-0.4784$ & $-0.3051_{-0.35}^{+0.35}$ & $-1.013$ & $0.4046$ \\
$b^{(2)}_{1 }$ &$2.478$ & $2.549_{-0.13}^{+0.13}$ & $2.289$ & $2.815$ &$2.525$ & $2.548_{-0.13}^{+0.13}$ & $2.294$ & $2.806$ \\
$b^{(2)}_{2 }$ &$0.2456$ & $0.3383_{-0.81}^{+0.75}$ & $-1.227$ & $1.893$ &$0.3543$ & $0.4051_{-0.8}^{+0.75}$ & $-1.144$ & $1.968$ \\
$b^{(2)}_{{\mathcal{G}_2} }$ &$-0.2815$ & $-0.287_{-0.41}^{+0.4}$ & $-1.108$ & $0.5181$ &$-0.3399$ & $-0.2646_{-0.4}^{+0.4}$ & $-1.066$ & $0.5351$ \\
$b^{(3)}_{1 }$ &$2.17$ & $2.276_{-0.12}^{+0.12}$ & $2.039$ & $2.519$ &$2.222$ & $2.247_{-0.12}^{+0.11}$ & $2.016$ & $2.48$ \\
$b^{(3)}_{2 }$ &$0.05868$ & $0.2079_{-0.64}^{+0.59}$ & $-1.026$ & $1.445$  &$0.495$ & $0.1881_{-0.64}^{+0.57}$ & $-1.012$ & $1.417$ \\
$b^{(3)}_{{\mathcal{G}_2} }$ &$-0.3944$ & $-0.4344_{-0.32}^{+0.32}$ & $-1.072$ & $0.2184$ &$-0.2487$ & $-0.3771_{-0.32}^{+0.32}$ & $-1.005$ & $0.2594$ \\
$b^{(4)}_{1 }$ &$2.247$ & $2.312_{-0.13}^{+0.12}$ & $2.064$ & $2.567$ &$2.254$ & $2.291_{-0.12}^{+0.12}$ & $2.049$ & $2.536$ \\
$b^{(4)}_{2 }$ &$0.4349$ & $0.01968_{-0.7}^{+0.65}$ & $-1.305$ & $1.399$ &$-0.125$ & $0.1048_{-0.71}^{+0.64}$ & $-1.226$ & $1.457$ \\
$b^{(4)}_{{\mathcal{G}_2} }$ &$0.02486$ & $-0.3231_{-0.37}^{+0.37}$ & $-1.06$ & $0.415$ &$-0.2762$ & $-0.3677_{-0.36}^{+0.36}$ & $-1.092$ & $0.3556$  \\\hline
$\Omega_{m }$ &$0.3319$ & $0.3388_{-0.018}^{+0.016}$ & $0.3039$ & $0.3736$ &$0.3412$ & $0.3493_{-0.018}^{+0.016}$ & $0.3159$ & $0.3832$ \\
$H_0$ & $68.96$ & $69.28_{-1.2}^{+1.1}$ & $67$ & $71.59$ &$69.23$ & $69.19_{-1.1}^{+1}$ & $67.04$ & $71.39$ \\
$\sigma_8$ &$0.7137$ & $0.6909_{-0.04}^{+0.036}$ & $0.6158$ & $0.7686$ &$0.7055$ & $0.7044_{-0.04}^{+0.035}$ & $0.6303$ & $0.7797$ \\
\hline
 \end{tabular}
 \caption{Full parameter constraints from the $\Lambda$CDM analysis of BOSS DR12 data using the power spectrum + bispectrum monopole datasets ($P_\ell$+$Q_0$+BAO + $B_0$, left) and including the bispectrum multipoles ($P_\ell$+$Q_0$+BAO + $B_\ell$, right). We give the best-fit values, the mean, 68\%, and 95\% confidence level results in each case, and show the derived parameters in the bottom rows. The superscripts on bias parameters indicate the sample, in the order NGC z3, SGC z3, NGC z1, SGC z1. Corresponding results with a \textit{Planck} prior on $n_s$ are shown in Tab.\,\ref{tab: all-constraints-fix-ns}.}\label{tab: all-constraints-free-ns}
\end{table}

 \begin{table}
    \centering
   \rowcolors{2}{white}{vlightgray}
  \scriptsize
     \begin{tabular}{|c|cccc|cccc|}
  \hline
  & \multicolumn{4}{c|}{$\bm{P_\ell + Q_0 + \mathrm{BAO} + B_0}$} & \multicolumn{4}{c|}{$\bm{P_\ell + Q_0 + \mathrm{BAO} + B_\ell}$}\\\hline
  \quad \textbf{Parameter}\quad\quad & best-fit & mean$\,\pm\,\sigma$ & \quad 95\% lower \quad & \quad 95\% upper \quad & best-fit & mean$\,\pm\,\sigma$ & \quad 95\% lower \quad & \quad 95\% upper \quad \\ \hline
$\omega_{\rm cdm }$ &$0.1242$ & $0.1262_{-0.0059}^{+0.0053}$ & $0.1152$ & $0.1374$ &$0.1284$ & $0.1302_{-0.0058}^{+0.0055}$ & $0.119$ & $0.1416$ \\
$h$ &$0.6809$ & $0.6832_{-0.0086}^{+0.0083}$ & $0.6665$ & $0.7002$ &$0.683$ & $0.6819_{-0.0081}^{+0.0078}$ & $0.6661$ & $0.6979$  \\
$ln\left(10^{10}A_{s }\right)$ &$2.69$ & $2.598_{-0.14}^{+0.13}$ & $2.335$ & $2.868$ & $2.626$ & $2.602_{-0.13}^{+0.12}$ & $2.354$ & $2.858$\\
$b^{(1)}_{1 }$ &$2.335$ & $2.366_{-0.13}^{+0.13}$ & $2.116$ & $2.619$ &$2.336$ & $2.372_{-0.12}^{+0.12}$ & $2.134$ & $2.616$ \\
$b^{(1)}_{2 }$ &$0.02127$ & $0.4025_{-0.78}^{+0.71}$ & $-1.08$ & $1.894$& $0.1334$ & $0.3783_{-0.76}^{+0.69}$ & $-1.046$ & $1.839$\\
$b^{(1)}_{{\mathcal{G}_2} }$ &$-0.08666$ & $-0.1787_{-0.34}^{+0.34}$ & $-0.8542$ & $0.5114$ &$-0.1642$ & $-0.1746_{-0.33}^{+0.34}$ & $-0.8431$ & $0.4913$ \\
$b^{(2)}_{1 }$ &$2.471$ & $2.502_{-0.13}^{+0.13}$ & $2.244$ & $2.765$ &$2.479$ & $2.511_{-0.13}^{+0.13}$ & $2.258$ & $2.765$ \\
$b^{(2)}_{2 }$ &$0.5985$ & $0.3585_{-0.78}^{+0.74}$ & $-1.155$ & $1.883$ &$-0.02707$ & $0.4099_{-0.81}^{+0.72}$ & $-1.07$ & $1.961$ \\
$b^{(2)}_{{\mathcal{G}_2} }$ &$-0.2001$ & $-0.1394_{-0.4}^{+0.39}$ & $-0.9225$ & $0.652$ &$-0.202$ & $-0.1906_{-0.38}^{+0.38}$ & $-0.9664$ & $0.5752$ \\
$b^{(3)}_{1 }$ &$2.17$ & $2.276_{-0.12}^{+0.12}$ & $2.039$ & $2.519$  &$2.222$ & $2.247_{-0.12}^{+0.11}$ & $2.016$ & $2.48$ \\
$b^{(3)}_{2 }$ &$0.2993$ & $0.2288_{-0.64}^{+0.58}$ & $-0.9686$ & $1.467$ &$0.01689$ & $0.2231_{-0.62}^{+0.56}$ & $-0.9552$ & $1.427$ \\
$b^{(3)}_{{\mathcal{G}_2} }$ &$-0.335$ & $-0.3312_{-0.32}^{+0.31}$ & $-0.9531$ & $0.2936$ &$-0.2906$ & $-0.2936_{-0.31}^{+0.31}$ & $-0.9069$ & $0.319$ \\
$b^{(4)}_{1 }$ &$2.207$ & $2.264_{-0.13}^{+0.12}$ & $2.022$ & $2.509$ & $2.239$ & $2.25_{-0.12}^{+0.12}$ & $2.014$ & $2.489$ \\
$b^{(4)}_{2 }$ &$-0.4185$ & $0.007792_{-0.7}^{+0.64}$ & $-1.296$ & $1.348$ &$0.3194$ & $0.1111_{-0.7}^{+0.62}$ & $-1.178$ & $1.46$ \\
$b^{(4)}_{{\mathcal{G}_2} }$ &$-0.3927$ & $-0.2497_{-0.36}^{+0.35}$ & $-0.9527$ & $0.4647$ &$-0.173$ & $-0.3235_{-0.35}^{+0.35}$ & $-1.023$ & $0.3781$ \\\hline
$\Omega_{m }$ &$0.3176$ & $0.3197_{-0.01}^{+0.0095}$ & $0.3004$ & $0.3393$ &$0.3246$ & $0.3295_{-0.01}^{+0.0096}$ & $0.31$ & $0.3491$ \\
$H_0$ &$68.09$ & $68.32_{-0.86}^{+0.83}$ & $66.65$ & $70.02$ 
& $68.30$
&$68.19_{-0.81}^{+0.78}$ & $66.61$ & $69.79$ \\
$\sigma_8$ &$0.7248$ & $0.7221_{-0.037}^{+0.032}$ & $0.6539$ & $0.7917$ &$0.739$ & $0.7356_{-0.035}^{+0.033}$ & $0.6701$ & $0.804$ \\\hline
 \end{tabular}
 \caption{As Tab.\,\ref{tab: all-constraints-free-ns}, but including a \textit{Planck} prior on the spectral slope $n_s$.}\label{tab: all-constraints-fix-ns}
\end{table}

\bibliography{short}
\bibliographystyle{JHEP}

\end{document}